\documentclass[%
 reprint,
superscriptaddress,
 amsmath,amssymb,
 aps,
]{revtex4-2}

\usepackage{graphicx}
\usepackage{dcolumn}
\usepackage{bm}
\usepackage[colorlinks=true, allcolors=blue]{hyperref}
\usepackage{amsmath}
\usepackage{amssymb}
\usepackage{physics}
\usepackage[capitalise]{cleveref}
\usepackage{dsfont}
\usepackage{subcaption}

\usepackage{ulem}
\usepackage[dvipsnames]{xcolor}
\usepackage[colorinlistoftodos,prependcaption,textsize=tiny]{todonotes}

\newcommand{\meanval}[1]{\langle#1\rangle}

\definecolor{colorMax}{HTML}{ff3300}

\usepackage[dvipsnames]{xcolor}
\definecolor{colorCosta}{HTML}{00b300}

\begin{document}

\preprint{APS/123-QED}

\title{A Passive and Self-Characterizing Cross-Encoded Receiver for Reference-Frame-Independent Quantum Key Distribution}

\author{Massimo Giacomin}
\email{massimo.giacomin.1@phd.unipd.it}
\affiliation{Department of Information Engineering, Università degli Studi di Padova}

\author{Francesco B. L. Santagiustina}
\altaffiliation[Currently at ]{ThinkQuantum s.r.l.}
\affiliation{Department of Information Engineering, Università degli Studi di Padova}

\author{Giuseppe Vallone}
\affiliation{Department of Information Engineering, Università degli Studi di Padova}

\author{Paolo Villoresi}\affiliation{Department of Information Engineering, Università degli Studi di Padova}

\author{Costantino Agnesi}
\email{costantino.agnesi@unipd.it}
\affiliation{Department of Information Engineering, Università degli Studi di Padova}

\date{\today}

\begin{abstract}
Quantum Key Distribution (QKD) promises to revolutionize the field of security in communication, with applications ranging from state secrets to personal data, making it a key player in the
ongoing battle against cyber threats. Reference-Frame-Independent (RFI) QKD aims to simplify QKD implementations by allowing to reduce the requirements of alignment on a shared reference frame. This is done by performing two mutually unbiased measurements on the control states. In this work, we present a novel fully passive receiver for time-bin encoded RFI-QKD. Conversion of time-bin to polarization is employed to perform the required quantum measurement in a fully passive manner. Furthermore, to overcome experimental errors, we retrieved a complete description of our measurement apparatus by employing a recently introduced Quantum Detector Self-Characterization technique, without performing tomographic studies on the detection stage. In fact, the security analysis carried out in this work uses experimentally retrieved Positive Operator Valued Measurements, which consider our receiver defects, substituting the ideal expected operators and thus increasing the overall level of secrecy. Lastly, we conducted a proof-of-principle experiment that validated the feasibility of our method and its applicability to QKD applications.
\end{abstract}

\maketitle

\section{Introduction}\label{sec:intro}
Quantum Key Distribution (QKD) is a cutting-edge technique that is transforming cryptography and cybersecurity in modern communication systems. It is a method that utilizes the fundamental postulates of quantum mechanics to generate and distribute cryptographic keys between two parties~\cite{BB84}. This innovative approach ensures an unparalleled level of unconditional security, which is crucial nowadays where the potential computational ability offered by quantum computers continues to increase~\cite{Shor1997, QC_Bouland2019, GoogleQC2019}. The essence of QKD lies in its unique ability to detect any form of eavesdropping, since any attempt to measure a quantum system invariably disturbs the system itself. 
As a result, keys are generated only when the information obtained by the eavesdropper is bounded below a threshold level that ensures that privacy amplification between legitimate users is possible, therefore guaranteeing the integrity of subsequent communications~\cite{Scarani2008}.\\
The choice of the most suitable encoding strategy is strongly impacted by the nature of the quantum channel along which the key is exchanged~\cite{Pirandola2019rev}. While polarization encoding is recognized for its reliability and minimal error rate~\cite{Agnesi:20}, making it well suited for free-space links, time-bin (TB) encoding~\cite{Bennett1992} is resistant to birefringence, making it an interesting solution for optical fiber networks. A relevant example is represented by aerial fiber links, usually integrated in sub-urban environments, where strong thermal and mechanical stresses may be present. Exposition to harsh external conditions causes the state of polarization (SOP) of light passing through them to dramatically fluctuate over time, worsening communication performance~\cite{Ding2017} and requiring complex polarization tracking methods to compensate~\cite{Li:18}. Furthermore, strong dispersion phenomena such as Polarization Mode Dispersion (PMD) are additional detrimental factors when the encryption strategy is polarization-based, while they are completely ineffective once the TB encoding is exploited. This holds in general, although recent studies have demonstrated the feasibility of polarization-based protocols in urban scenarios \cite{Avesani:21, Telebit}, which, however, require active polarization compensation within key exchange.

Despite the robustness against dispersive phenomena shown by TB encoding, protocols based on this technology demand precise control in the stabilization of the interferometric measurements performed both on the transmitter and receiver sides~\cite{Makarov:04,Svarc:23, Hacker_2023}. A suboptimal realization may result in the degradation of the final quantum bit error rate (QBER), which impacts the overall efficiency of the cryptographic procedure.

To this end, interest has recently been shown in the family of Reference-Frame-Independent (RFI) protocols~\cite{originalRFI2010}, in which the requirement over the relative phase between the transmitter and receiver measurement bases can be dropped~\cite{previous_rfi_2,RFI_QKD_china, Wang2019, Chen2020, Tang2022, RFI_waterloo}. As a matter of fact, this alternative allows to assume a (slow enough) phase-drift between the sender and the receiver reference frames, reducing the complexity of the overall system since no active reference-frame calibration is needed. In this solution, three (or even two \cite{RFI_QKD_china}) mutually unbiased bases are required, in which at least one is used to monitor the eavesdropper (Eve)'s information collected during the key exchange and one basis is necessary to generate and distribute the final raw key. The only requirement imposed by this strategy is to have a stable and well-aligned key generation basis. This is the case of the TB encoding, since photons time-of-arrival (TOA) is naturally stable and independent of phase drifts of the transmitter or receiver interferometers. By mapping the early $\ket{\mathcal{E}}$ and late $\ket{\mathcal{L}}$ TOAs to the computational basis states $\ket{0}$ and $\ket{1}$, the choice of this stable basis falls on $\mathbb{Z}$ ($\hat{\sigma}_Z$ according to Pauli's notation), while the others, depending on the phase component, fluctuate in time, resulting in a quantum state that spans the equatorial $\mathbb{X}-\mathbb{Y}$ ($\hat{\sigma}_X-\hat{\sigma}_Y$) plane of the Bloch sphere.

Up to now, all prepare-and-measure time-bin encoded RFI-QKD demonstrators have relied on active receivers where the decoding interferometer is equipped with a fast phase modulator that randomly imposes a $0$  or $\pi/2$ phase shift~\cite{RFI_QKD_china,Wang2019, Chen2020, Tang2022}. This active approach is required to guarantee the measurement in two mutually unbiased bases that lay on the equatorial plane of the Bloch sphere. Although this approach is valid and secure, it comes with several implementation drawbacks. First, it necessitates real-time synchronization between the transmitter and receiver with stringent requirements on frequency drifts and jitters. This is because the random phase shift at the receiver should be well-centered on the incoming optical signal to ensure an effective modulation and prevent crosstalk between adjacent qubits. Secondly, to ensure maximum protocol security, the random modulation should be determined by the output of a secure entropy source such as a Quantum Random Number Generator (QRNG)~\cite{Jennewein2000, Stipcevic2007, QRNG2020}. This adds additional costs to the receiver and substantially increases the architectural complexity of the system. In fact, this requires the development of custom high-speed electronics that connect to the QRNG, handle its bit stream to select the basis, save the selection in memory for the sifting and post-processing stages of the QKD protocol, and ultimately produce the synchronized electrical signals needed to create the phase shift in the receiver's interferometer~\cite{Stanco2022}. Another disadvantage of this active method is the insertion loss caused by the fast phase modulator. These modulators are typically made using Lithium Niobate crystals, which inherently have around 3dB loss.

To overcome these drawbacks, here we introduce a novel receiver design for time-bin encoded RFI-QKD that implements the required measurements prescribed by the protocol in a fully passive manner. This is achieved by employing the time-bin to polarization conversion introduced by our research group in \cite{cross-encoding}. This cross-encoding allows us to exploit the robustness and stability of polarization optics to passively implement the two mutually unbiased bases. To the best of our knowledge, this is the first proposal for a fully-passive decoder for prepare-and-measure time-bin encoded RFI-QKD and can represent a substantial increase in technological maturity for RFI-QKD since it considerably reduces implementation and deployment complexity.

A further innovation presented in this work is the use of a realistic description of the measurement apparatus for the security analysis of the RFI-QKD protocol. This description of the utilized receiver was possible employing the Positive-Operator-Valued-Measurements (POVMs) formalism that can take into account detrimental factors and defects of the real physical apparatus. Some examples of these defects are  misaligned waveplates, non-ideal BS and PBS, non-zero surface reflections, lossy fiber couplings, non-uniform single-photon detection efficiencies to name but a few. In order to overcome idealized device descriptions and adopt a realistic POVM approach, we developed a method that relies on the innovative Quantum Detector Self-Characterization (QDSC) technique introduced by A. Zhang \textit{et al.}~\cite{QDSC}. With this procedure it is possible to gain a complete description of the setup, including the inevitable defects typical of real apparati, without the requirement of performing a complete tomographic studies on the detection stage. With the resulting POVMs, the security analysis carried out in this work utilizes a realistic descriptions of the receiver, and allows us to substitute the ideal operators conventionally used in these types of analysis. This, therefore, increases the overall level of secrecy since no assumptions are made about the receiver features, except its ability to measure 2-dimensional objects, i.e., qubits. As far as we are aware, this work represents the first use of QDSC for the evaluation of the security of a QKD protocol, which can lead to more robust security analysis and even higher security levels for quantum cryptography.

In the following sections we will describe the hardware and the methods behind our passive and self-characterizing cross-encoded receiver for RFI-QKD, followed by a proof-of-concept experimental implementation.  

\section{Methods}\label{sec:methods}
We will first begin with a hardware description of the cross-encoded receiver for the RFI-QKD. Then we will describe the QSDC approach to obtain POVMs that provide a realistic description of the receiver. Lastly, we will describe the security analysis of the RFI-QKD protocol, where ideal detector descriptions are replaced with the realistic POVMs previously obtained.

\subsection{Measurement Apparatus Description}\label{sec:hwDescription}
\begin{figure*}
    \centering
    \includegraphics[width=0.88\linewidth]{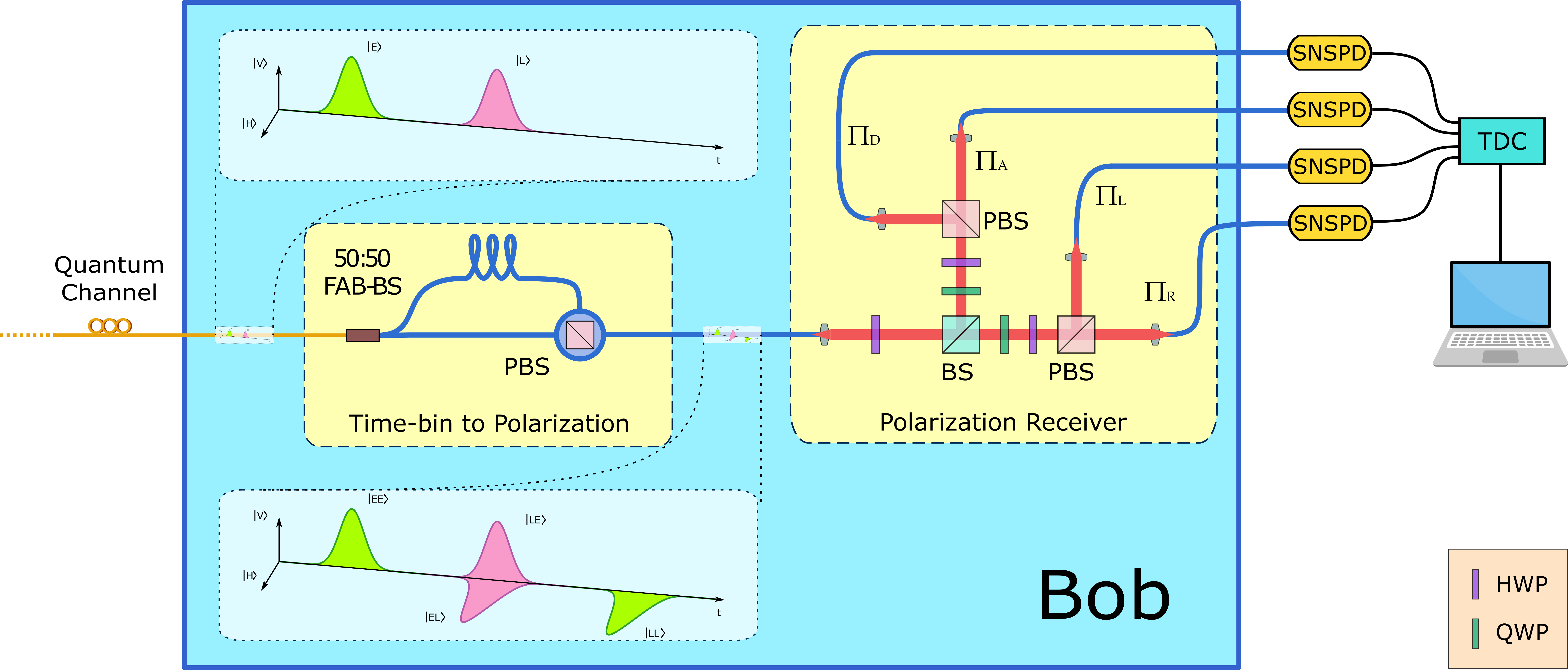}
    \caption{Schematic representation of the built receiver (Bob), composed by the TB-to-polarization converter, the polarization receiver and the TOA detection stage. BS: Beam Splitter, PBS: Polarizing BS, HWP: Half Wave Plate, SPD: Single Photon Detector, TDC: Time-to-Digital Converter. Light arriving with TB encoding along the QC is distributed into two-peaks, enters the FAB-BS and is split whether in the long or short path. The two recombine at the PBS forming the typical three-peaks pulse before being injected into the polarization receiver. The HWPs and QWPs at the outputs of the free-space BS are set such that the polarization receiver performs measurements in the $\mathbb{X}$ ($\mathbb{Y}$) basis after being reflected (transmitted).}
    \label{fig:rx_scheme}
\end{figure*}
The key object here is the time-bin to polarization converter (shown in \cref{fig:rx_scheme}), whose aim is to directly map the information encoded in the TOA of the photons into their polarization degree of freedom. This conversion is ensured by an \textit{Unbalanced-Mach-Zender-Interferometer} (UMZI) which begins with a \textit{Fast-Axis-Blocking} (FAB) Beam Splitter (BS) that randomly routes the incoming photons in the short or long paths. The photons then recombine in PBS such that the UMZI outputs horizontal or vertical SOPs depending on the arm walked by the light. The result of the combination of Bob's UMZI with the Alice TB encoding temporally distributes the light in the three-peak configuration, typical of TB realizations, although each peak has its own distinct polarization. 

In fact, the relation between the time-bin encoding and the polarization encoding follows the mapping
\begin{equation}
\begin{split}
\alpha\ket{\mathcal{E}} +& \beta\ket{\mathcal{L}} \\ \Downarrow & \\   \frac{1}{\sqrt{2}} \Bigl( \alpha\ket{\mathcal{EE}}\otimes\ket{H} +& e^{i\phi_\mathrm{B}} \alpha \ket{\mathcal{EL}}\otimes\ket{V} + \\ +  \beta\ket{\mathcal{LE}}\otimes\ket{H} &+  e^{i\phi_\mathrm{B}}\beta \ket{\mathcal{LL}}\otimes\ket{V} \Bigr)
\end{split}
\end{equation}
where $\phi_\mathrm{B}$ is the intrinsic phase of Bob's UMZI.
In particular, when Alice sends $\ket{+} = (\ket{\mathcal{E}}+\ket{\mathcal{L}})/\sqrt{2}$, the output state is
\begin{equation} \label{eq:superpos_EL}
\begin{split}
\ket{\Psi_+} =  \frac{1}{2} \Bigl( & \ket{\mathcal{EE}}\otimes\ket{H} + \\ & + e^{i\phi_\mathrm{B}} \ket{\mathcal{EL}}\otimes\ket{V} + \ket{\mathcal{LE}}\otimes\ket{H} + \\
                                & +  e^{i \phi_\mathrm{B}} \ket{\mathcal{LL}}\otimes\ket{V} \Bigr).
\end{split}
\end{equation}
 
It is important to note that the lateral peaks $\ket{\mathcal{EE}}$ and $\ket{\mathcal{LL}}$ correspond to light traveling along the short or long arms of both transmitter and receiver's UMZI and since those TOAs are a measurement in the $\mathbb{Z}$ basis, they are used to generate the secret key. Given that 50\% of the light falls in these lateral peaks, this is not a negligible contribution to the final key rate.
However, only the central peak contains the superposition between the indistinguishable early-late $\ket{\mathcal{EL}}$ and late-early $ \ket{\mathcal{LE}}$ components, and the relative phase information between them is encoded in the polarization state $\ket{\Psi_{c}} = (\ket{H}+e^{-i\phi_B}\ket{V})/\sqrt{2}$ of the light, which lies on the equatorial plane of the Bloch Sphere defined by the $\mathbb{X} = \{ \ket{D} = (\ket{H}+\ket{V})/\sqrt{2}, \ket{A} = (\ket{H}-\ket{V})/\sqrt{2}\}$ and $\mathbb{Y} = \{ \ket{L} = (\ket{H}+i\ket{V})/\sqrt{2}, \ket{R} = (\ket{H}-i\ket{V})/\sqrt{2}\}$ bases. The central peak is therefore exploited to implement the phase error determination.

Since the proposed protocol fulfills the RFI hypothesis of having the key generation basis in a stable reference frame for both Alice and Bob, whereas the control states slowly drift in a confined qubit subspace, it is legit to assume that this phase error estimation can be performed by measuring the received control states in the equatorial plane of the Bloch sphere.
The scheme of the passive receiver built to carry out this step is depicted in \cref{fig:rx_scheme}. The polarization decoder is based on a standard and well-validated design which is often exploited for polarization-encoded BB84 experiments such as~\cite{Liao2017_Sat,Berra2023}.

Here, further considerations can be made. At the entrance of the polarization receiver, the photon has the same probability $(\frac{1}{2})$ to be transmitted or reflected during the passage across the BS. This represents the purely random selection performed by Bob in the choice of which basis to use for the detection of Eve presence in the channel. Depending on which BS output the photon exits from, its SOP is rotated by properly tuned HWP and QWP in order to implement the projective measurements on $\mathbb{X}$ and $\mathbb{Y}$. At this point the interaction with the respective PBS routes the photon toward the proper detector arm, and here its time of arrival is registered by a TDC.

\subsection{POVM determination via QDSC}\label{sec:qdsc}
In Quantum Mechanics, the measurement procedure inherently prevents the complete characterization of a quantum state through tomographic reconstruction without the availability of precisely calibrated probe states. Furthermore, calibrating the probe states source depends on the precision of the measurement apparatus used in the calibration, creating a circular paradox. To address this issue, self-characterizing methods, such as the QDSC introduced by A. Zhang \textit{et al.}~\cite{QDSC} can be employed. Here we specialize that method for the POVM determination of a cross-encoded receiver for time-bin encoding prepare-and-measure RFI-QKD.  

The core idea behind QDSC is to characterize general unknown quantum measurements exploiting solely the detector outcomes of the device itself from random and uncharacterized input states, in order to retrieve the accessible region of outcomes at the disposal of the measurement device, named \textit{response range} and described formally as follows:
\begin{equation}
    \mathcal{W} := \{[\text{Tr}(\rho\Pi_0),\dots,\text{Tr}(\rho\Pi_{n-1})]|\rho\geq0,\text{Tr}(\rho)=1\}.
\end{equation}
As noted by M. Dall'Arno~\textit{et al.}~\cite{DallArno2017}, in the case of qubits, the response range $\mathcal{W}$ corresponds to a hyper-ellipsoid lying in an \textbf{n}-dimensional space. This hyper-ellipsoid is determined by the matrix $Q$  and centered in \textbf{t} according to the following equation
\begin{equation}\label{eq:POVMsDecomposition}
    (\textbf{p} - \textbf{t})^T Q^+ (\textbf{p} - \textbf{t}) \leq 1.
\end{equation}
The matrix $Q$ and vector \textbf{t} are connected to the POVMs $\pi_k$ via the definitions 
\begin{equation}
    \begin{cases}\label{eq:POVMsDescription}
        Q_{k,h} = \textbf{m}_k^T \textbf{m}_h = \frac{1}{2}Tr(\Pi_k \Pi_h) - \frac{1}{4}Tr(\Pi_k)Tr(\Pi_h)\\
        t_k = \frac{1}{2}Tr(\Pi_k).
    \end{cases}
\end{equation}
The matrix $Q$ quantifies the overlap between POVMs elements, whereas \textbf{t} represents the weight vector of POVMs.
The explicit derivation of these relations can be found in \cref{sec:appendixA}.

The complexity of the problem is thus reduced to the determination of the hyper-ellipsoidal response range $\mathcal{W}$ that best fits the statistics of the measurement outcomes $\{p^{(j)}\}$.
This can be mapped to an optimization problem, according to the following:
\begin{equation}
    \begin{aligned}\label{eq:opt_problem}
    \texttt{minimize} &: \quad \sum_{j \in \mathcal{B}} \Bigl[1 - (p^{(j)} - t)^T \cdot Q^+ \cdot (p^{(j)} - t) \Bigr]^2 \\
    \texttt{subject to } &: \quad t_k^2 - Q_{k,k} \geq 0.
    \end{aligned}
\end{equation}
From the estimation of the response range it is then possible to retrieve the nature of the POVMs involved in the measurement process by inverting \cref{eq:POVMsDescription}.

Our experimental procedure starts with the collection of single counts at the output of the passive receiver. The data collected from a TDC consist of four sets of counts rates, associated with the receiver outputs and TOAs, that change in time as the interferometric phase drifts due to environmental conditions.\\ Each counts rate detected from a specific receiver channel can be normalized with respect to all four counts rates registered in the same TOA (here we only use the central peak), thus obtaining the experimental frequencies that are normalized according to $\sum_k p_k^{(j)} = 1$ for each $j$. These are formally equivalent to the probabilities of detecting a specific polarization in each of the receiver outputs. Assuming a \textit{4}-outcome measurement process probed with \textit{m} random quantum states, the collected statistics can be described as
\begin{equation}
    P_{4 \times m} = \begin{pmatrix}
        p_1^{(0)} & p_1^{(1)} & \dots & p_1^{(m-1)} \\
        p_2^{(0)} & p_2^{(1)} & \dots & p_2^{(m-1)} \\
        p_3^{(0)} & p_3^{(1)} & \dots & p_3^{(m-1)} \\
        p_4^{(0)} & p_4^{(1)} & \dots & p_4^{(m-1)} \\
    \end{pmatrix}
\end{equation}
where each element of the matrix is formally described by the Born's rule according to
\begin{equation}
    p_k^{(j)} = Tr(\rho^{(j)}\Pi_k).
\end{equation}
Here, $\{\rho^{(j)}\}$ represents a quantum state in the density matrix representation and $\{\pi_k\}$ is the \textit{k}-th POVM, with $k = \{L, R, D, A\}$. In particular, in our experiment $\rho^{(j)}$ is randomly distributed over the equatorial plane of the Bloch sphere due to the phase drift induced by thermal and mechanical stresses from the environment. 

According to the analysis proposed by A. Zhang \textit{et al.}~\cite{QDSC}, thanks to the linear dependencies of the measurement operators, the dimension of both the collected statistics $P$ and the response range $\mathcal{W}$ can be reduced to a \textbf{3}-dimensional space. However, since our protocol exploits only two control bases, the dimension can be further reduced to \textbf{2}, thus lowering the complexity of the considered problem. This dimension reduction process is performed using a Principal Component Analysis (PCA)~\cite{doi:10.1098/rsta.2015.0202} and is further described in \cref{sec:appendixA}. The outcome of this process leads to the reduced matrix $\Tilde{A}_{3 \times m}$ whose each column element can be interpreted as the spatial representation of the collected states in the reduced probability space. It is, however, important to notice that in our case the elements of the third row are all approximately zero, confirming that all data lie in a 2-dimensional subspace.  

Given that the response range $\mathcal{W}(\pi)$ is a convex set, and therefore each inner point can be obtained by means of a linear combination of the external boundary coordinates, we are only interested in the boundary data of the whole collection in order to describe it. This is obtained by means of a Convex-Hull Boundary filtering. With these filtered data, it is then possible to perform a direct ellipse fit, which can be mapped to the optimization problem stated in \cref{eq:opt_problem}, therefore obtaining the matrix $Q$ and the vector $t$ that characterize the POVM elements.

\subsection{Security Analysis}\label{sec:security}
Once the physical POVMs have been derived experimentally, the RFI procedure of the protocol can be carried out.\\
The final goal in the implementation of a QKD realization is for sure to guarantee a sufficient secret key rate in order to allow two users to exchange enough cryptographic material for their private communication.

The generic description of the secret key rate fraction, usually expressing the amount of secure information that can be extracted from a specific protocol, considering any possible strategy attack adopted by Eve, is formally outlined by
\begin{equation}
    R = 1-h\bigl(e_{\hat{\mathbb{Z}}\hat{\mathbb{Z}}}\bigr)-I_E.
\end{equation}
Here, the term $I_E$ estimates the information acquired by Eve during the exchange of the raw key. In the assumption of maintaining the QBER under the upper bound $e_{\hat{\mathbb{Z}}\hat{\mathbb{Z}}}~\le~ 15.9\%$~\cite{RFI_QKD_china}, this leakage of secrecy can be computed as
\begin{equation}\label{eq:Ie}
    I_E=(1-e_{\hat{\mathbb{Z}}\hat{\mathbb{Z}}})\cdot h \biggl(\frac{1+\mu}{2}\biggr)+e_{\hat{\mathbb{Z}}\hat{\mathbb{Z}}}\cdot h \biggl(\frac{1+\nu(\mu)}{2}\biggr)
\end{equation}
in which $h(x)$ denotes the \textit{binary Shannon entropy}
\begin{equation}
    h(x,\bar{x})= - P(x)\cdot log_2 \bigl[P(x)\bigr] - P(\bar{x})\cdot log_2 \bigl[P(\bar{x})\bigr],
\end{equation}
and the parameters $\mu$ and $\nu$ are expressed as
\begin{align}
    &\mu=min\biggl[\frac{\sqrt{C/2}}{1-e_{\hat{\mathbb{Z}}\hat{\mathbb{Z}}}},1\biggr] \\ &\nu=\frac{\sqrt{C/2-\bigl(1-ee_{\hat{\mathbb{Z}}\hat{\mathbb{Z}}}\bigr)^2\mu^2}}{e_{\hat{\mathbb{Z}}\hat{\mathbb{Z}}}}.
\end{align}
The correlation parameter C, according to the study proposed in the last years~\cite{previous_rfi_1,previous_rfi_2,RFI_QKD_china}, is the most useful parameter that evaluates the level of security in a RFI protocol. It is a measure of the correlation between the information possessed by Alice and Bob, inversely proportional to the information gathered by Eve, and it is described as
\begin{equation}\label{eq:C_param}
C=\meanval{\hat{\mathbb{X}}_A\hat{\mathbb{X}}_B}^2 + \meanval{\hat{\mathbb{X}}_A\hat{\mathbb{Y}}_B}^2 + \meanval{\hat{\mathbb{Y}}_A\hat{\mathbb{X}}_B}^2 + \meanval{\hat{\mathbb{Y}}_A\hat{\mathbb{Y}}_B}^2,
\end{equation}
where the notation assumes $\{\hat{\mathbb{X}},\hat{\mathbb{Y}},\hat{\mathbb{Z}}\} \equiv \{\hat{\sigma}_X, \hat{\sigma}_Y, \hat{\sigma}_Z\}$. Furthermore, the final QBER can be evaluated as
\begin{equation}\label{eq:qber}
    QBER=e_{\hat{\mathbb{Z}}\hat{\mathbb{Z}}}=\frac{1-\meanval{\hat{\mathbb{Z}}_A \hat{\mathbb{Z}}_B}}{2}.
\end{equation}
The only two conditions imposed here are to have a well define direction, that is $\hat{\mathbb{Z}}_A=\hat{\mathbb{Z}}_B$, and to have the other two directions to slowly vary in time, according to the following
\begin{equation}
    \begin{aligned}
    \hat{\mathbb{X}}_B&=cos(\beta) \hat{\mathbb{X}}_A+sin(\beta) \hat{\mathbb{Y}}_A\\ \hat{\mathbb{Y}}_B&=cos(\beta) \hat{\mathbb{Y}}_A-sin(\beta) \hat{\mathbb{X}}_A.
    \end{aligned}
\end{equation}
Theoretically, the maximum value achievable in Eq.~\ref{eq:C_param} is $C=2$, under the condition of utilizing two maximally entangled states in the description of the two quantum states possessed by Alice and Bob after the distribution of a single bit of information. In this case, the parameter $e_{\hat{\mathbb{Z}}\hat{\mathbb{Z}}}$ is found to be zero. 

Finally, the target of this RFI-QKD protocol is to estimate a lower bound on the C parameter compatible with the experimental observations. Following the argument proposed in \cite{RFI_QKD_china}, this issue can be faced assuming it to be a \textit{minimization Semi-Definite Programming} (SDP) problem in the equivalent entanglement-based version of the protocol, which can be dealt imposing the following:
\begin{equation}\label{eq:cvx_problem}
    \begin{aligned}
    \underset{\hat{\rho}_{AB}}{\texttt{minimize}}\texttt{ : }& \quad C \\
    \texttt{subject to : }& \begin{cases} Tr\Bigl(\hat{E}_{ZZ}\hat{\rho}_{AB}\Bigr)=e_{\hat{\mathbb{Z}}\hat{\mathbb{Z}}}\\ Tr\Bigl(\hat{P}_{+}^A \otimes \hat{\Pi}_{\chi j}^B ~\hat{\rho}_{AB}\Bigr)=p_{+,\chi j} \\ Tr\bigl(\hat{\rho}_{AB}\bigr)=1 \\ \hat{\rho}_{AB} \geq 0 \end{cases}
    \end{aligned}
\end{equation}
where $\{\chi\} \in \{\mathbb{X},\mathbb{Y}\}$ are the possible bases to be chosen and $\{j\} \in \{0, 1\}$ the classical symbols encoded in the photons. Furthermore, the notations $\hat{E}_{ZZ}$ and $\hat{P}_i=\ket{i}\bra{i}$ indicate, respectively, the \textit{error operator} in the $\mathbb{Z}$ basis and a \textit{projective measurement} performed on the entangled state $\hat{\rho}_{AB}$. The symbol $\hat{\Pi}_{\chi j}^B$ represents each POVM reconstructed experimentally applying QDSC. The use of this realistic representation of the receiver apparatus represents our main contribution to the RFI-QKD security analysis. Finally, the term $p_{+,\chi j}$ stands for the experimental frequency that Bob measures with the POVM element $\Pi_j$ given that Alice has sent the state $\ket{+}$. As a matter of fact, the natural phase drift experienced by both Alice's (if present) and Bob's UMZI is the direct result of thermal and mechanical stresses experienced from the surrounding environment. Therefore, the quantum state $\ket{+}$ sent by Alice will lead to different $p_{+,\chi j}$ distributions over time.


Finally, a last comment must be made regarding the finite-key analysis performed in this research. According to \cite{Sheridan2010}, the generation of non-asymptotic keys by means of a RFI protocol imposes to assume a statistical deviation with respect to the infinite-key version of an amount of 
\begin{equation}
    \delta(k) = \sqrt{\frac{\ln(1/\epsilon)+2\ln(k+1)}{2k}}.
\end{equation}
where $k$ indicates the number of photons used in each run of the minimization in \cref{eq:cvx_problem}, and $\epsilon$ represents the security parameter, here assumed to be $10^{-5}$. This conservative term has been used to compute the upper and lower bounds in the minimization problem, replacing the equality sign "$=$" with "$\leq$" and "$\geq$", according to the following logic
\begin{equation*}
    x = y \longrightarrow \begin{cases}
        x \geq y - \delta(k) \\ x \leq y + \delta(k) 
    \end{cases},
\end{equation*}
where $x$ and $y$ represent two expressions of a generic equation.

The presented optimization problem can be solved exploiting the \texttt{MATLAB} package \texttt{CVX}, designed to perform convex optimization operations \cite{cvx_1,cvx_2}.

\section{Results}\label{sec:results}
 Since this work mainly concentrates on the innovative receiver design of time-bin encoded RFI-QKD, a simple and passive transmitter was implemented for the proof-of-principle demonstration of our proposal. The single photons exchanged from Alice to Bob were generated inside a three-stages transmitter, consisting of a pulsed laser source and a polarization to time-bin converter. We exploited a \texttt{Mira\textsuperscript{TM} HP-P Ti:Sapphire Laser} in \textit{mode-locking} configuration to generate optical pulses at $775$~nm with a repetition rate of $76$~MHz, gathering in a single mode fiber about $1$~mW of pulsed coherent light. This light was then used to pump a type-II PPKTP SPDC crystal that generated two photons at $1550$ nm with crossed polarization with an emission probability of around 10\%. The horizontally polarized photons proceed to the consecutive stage, whereas the vertically polarized ones are directly measured as a herald. At this point, single photons with horizontal polarization cross a $45^\circ$  tilted HWP before being injected into the encoding converter. This is realized with an UMZI, where the input element is constituted by a free-space PBS, which assigns two  paths of different lengths (path difference of $2.5$ ns) to the horizontal and vertical components of the input photons, which then recombine in a fiber FAB-BS before being launched into the QC. The FAB-BS erases all polarization information, ensuring a single polarization state at the output. Furthermore, one of the two branches of the UMZI has a tunable free-space delay line that allows to maximize the interference of the received photons with Bob's UMZI. The output state of this stage is described by
 \begin{equation}
     \ket{\Psi_{out}} = \frac{1}{\sqrt{2}}(\ket{\mathcal{E}}+e^{i\phi_A}\ket{\mathcal{L}})
 \end{equation}
where $\phi_A$ is a randomly fluctuating phase imparted by the encoding UMZI. The insertion loss of the transmitter is around 5 dB mainly due to the FAB-BS at the closure of the UMZI and non-optimal couplings into single mode fibers.

The QC in this experiment was represented by a $50$ km spool of standard single mode optical fiber (with losses of 0.2 dB/km, resulting in a total loss of about 10 dB)  connected with a polarization controller (PC), necessary to maximize the total counts rate. This procedure is required by our polarization sensitive receiver apparatus, and it is a fairly common situation for TB receivers observed in several experimental demonstrations~\cite{cross-encoding,RFI_QKD_china,Wang2019, Chen2020, Tang2022,Sasaki:11,Dynes:12,Dixon:15}.

As already mentioned in \cref{sec:hwDescription}, light passing through the QC is then collected by the receiver UMZI to be converted in polarization encoding and then measured in the proper detector. For this purpose, we exploited four Superconducting Nano-wire Single Photon Detectors (SNSPDs), developed by \texttt{ID Quantique}. This device is integrated in an automated closed-cycle $0.8$~K cryostat, capable of guaranteeing at least $80\%$ efficiency (at $\lambda=1550$~ nm). The main features are a \textit{timing jitter} at most $50$~ ps, a \textit{dark count rate} smaller than $100$~Hz and a \textit{recovery time} not exceeding $80$~ns. The conversion from an analog time of arrival to a digital datum is accomplished by a \texttt{quTAG} \textit{time-to-digital-converter} (TDC), powered by \texttt{qutools}. The insertion loss of the receiver was measured around 3 dB, including detector efficiency. Synchronization between the transmitter and the receiver was achieved following the algorithm introduced in \cite{Santagiustina:24}.

\begin{figure}
    \includegraphics[width=0.999\linewidth]{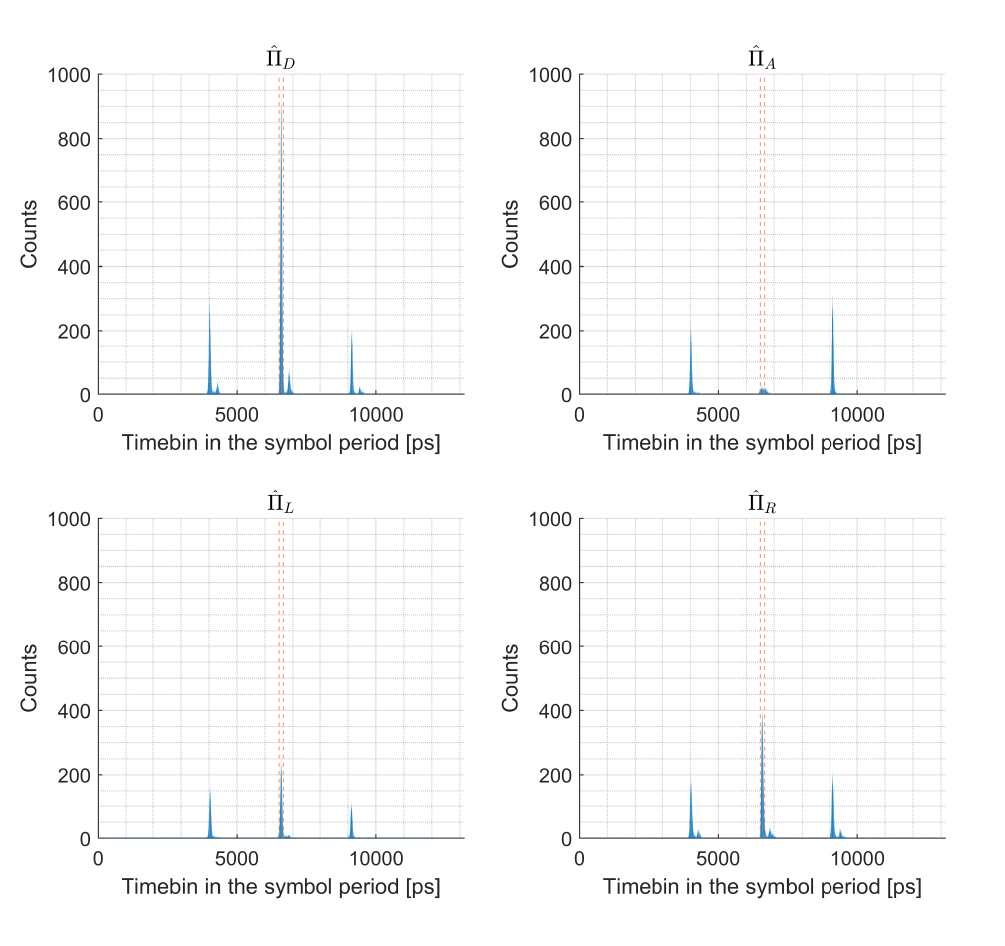}
    \caption{ The histograms corresponding to the counts collected at the four receiver outputs, together with the $150$~ps time window (red dashed vertical lines) associated with the central peak. Each histogram is obtained from $2$ seconds of integration.}
    \label{fig:TB}
\end{figure}

An average of about 120 kHz of detection is measured on all four channels and all TOAs. This therefore corresponds to 60 kHz in the central peak, which is useful for the security assessment of the QKD protocol. Considering the experimental data, a typical TB signal that is collected at the receiver side is depicted in \cref{fig:TB}. From each histogram, obtained from $2$-seconds of integration, and for each receiver output, we derived the total counts belonging to each TOA by choosing a temporal window of 150 ps. It is interesting to note that the 2 seconds selected for \cref{fig:TB} depicts the output statistics of a state close to the $\ket{D}$ polarization. In fact the diagonal (anti-diagonal) output port show constructive (destructive) interference, whereas the circular outputs are well balanced.

In the following subsections we will present the experimental results obtained following the analysis procedures described in the previous sections. This allowed us to determine the experimental POVMs associated with the receiver measurement, and subsequently calculate the associated security parameter and secure key rate fraction of our proof-of-principle experiment.

\subsection{Experimental POVM determination}
\begin{figure}
    \centering
    \includegraphics[width=0.92\linewidth]{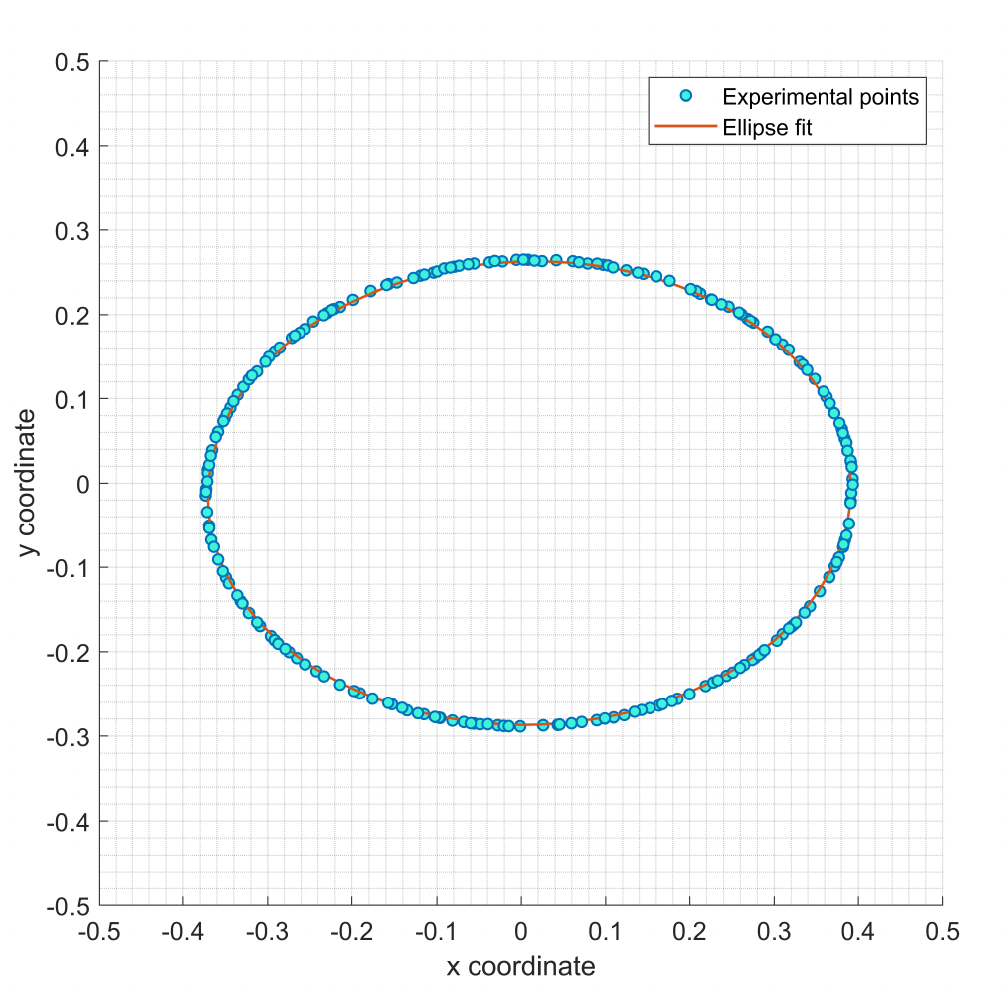}
    \caption{Response range of the characterized receiver. The experimental data are represented by the blue points, while the orange line describes the elliptical fit.}
    \label{fig:ellipseFit}
\end{figure}

\begin{figure}
    \centering
    \includegraphics[width=0.99\linewidth]{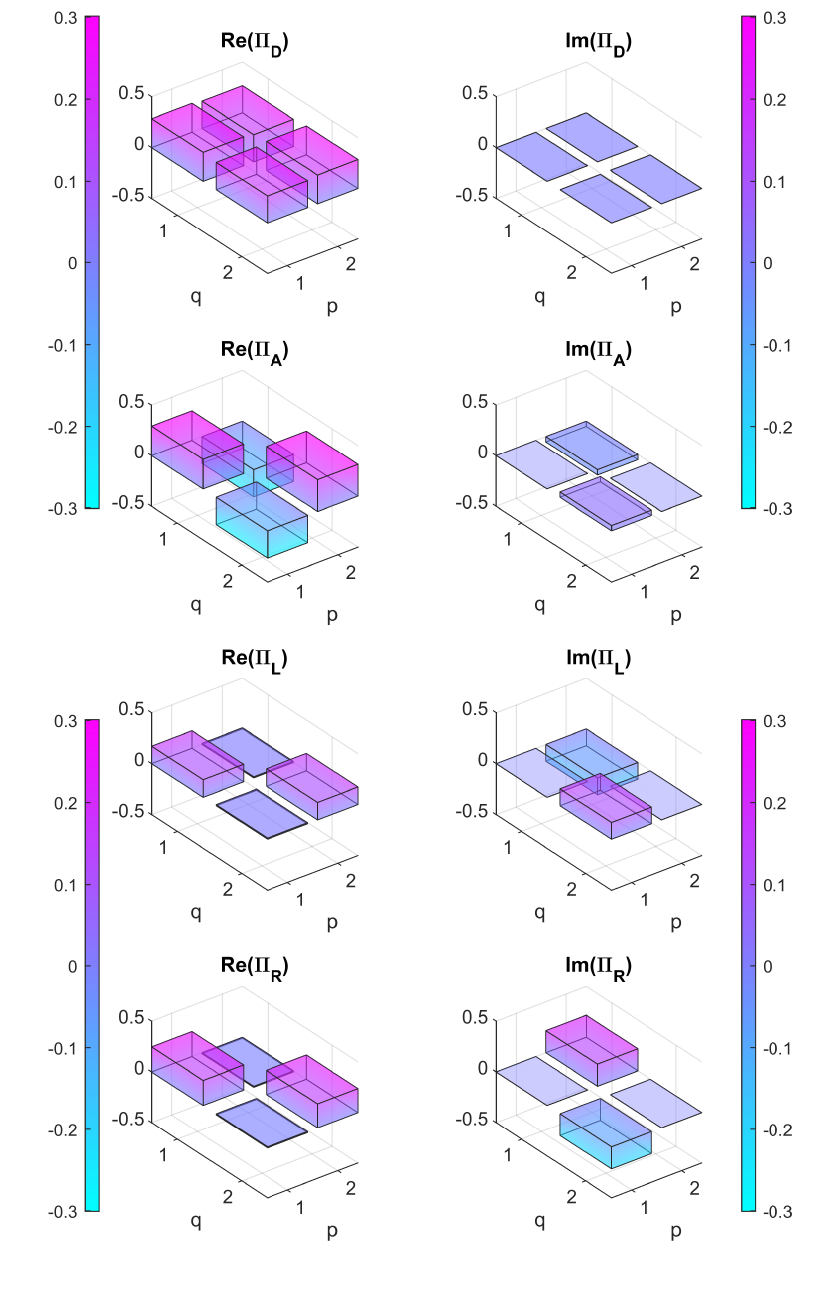}
    \caption{Graphic representation of the retrieved POVMs with their real part (\textit{left column}) and imaginary part (\textit{right column}).}
    \label{fig:povm_bars}
\end{figure}

By exploiting the natural drift of the encoder's output state and following the procedure introduced in \cref{sec:qdsc} we retrieved the response range of the implemented measurement apparatus. Considering our specific case of four measurements corresponding to two MUBs, the reduced probability space obtained after PCA consists of a plane, resulting in an elliptically shaped response range, as shown in \cref{fig:ellipseFit}.
From these selected boundary points, we then performed an elliptical fit, allowing us to calculate the POVM elements associated with our measurement apparatus by inverting \cref{eq:POVMsDescription}.
The retrieved experimental POVMs are shown in \cref{fig:povm_bars}. In \cref{sec:appendixB}, further experimental details are presented. 

It is interesting to compare the recovered POVMs with the idealized measurements $\{(\ketbra{D}{D})/2, (\ketbra{A}{A})/2, (\ketbra{L}{L})/2, (\ketbra{R}{R})/2\} $ where the factor $1/2$ comes from the ideal $50:50$ beam splitter at the beginning of the measurement apparatus. To perform this comparison we calculate the fidelity $\mathcal{F}$ according to the following formula:
\begin{equation}\label{eq:fidelity}
    \mathcal{F}(\Pi_i^\mathrm{id}, \Pi_i^\mathrm{exp})  = \frac{\mathrm{Tr}\left(\sqrt{\sqrt{\Pi_i^\mathrm{id}}\Pi_i^\mathrm{exp}\sqrt{\Pi_i^\mathrm{id}}}\right)^2}{ \mathrm{Tr}(\Pi_i^\mathrm{id}) \mathrm{Tr}(\Pi_i^\mathrm{exp})}
\end{equation}
where $\Pi_i^\mathrm{id}$ is the idealized measurement and $\Pi_i^\mathrm{exp}$ is the experimentally retrieved POVM. In \cref{tab:fidelity} we report the calculated fidelities, where a noticeable difference between real and idealized measurements can be observed. The main reason for this difference can be attributed to non-homogeneous optical losses between the optical branches in the receiver apparatus, different quantum efficiencies of the detectors and small defects in the polarization optics.
\begin{figure*}
    \centering
    \includegraphics[width=0.98\linewidth]{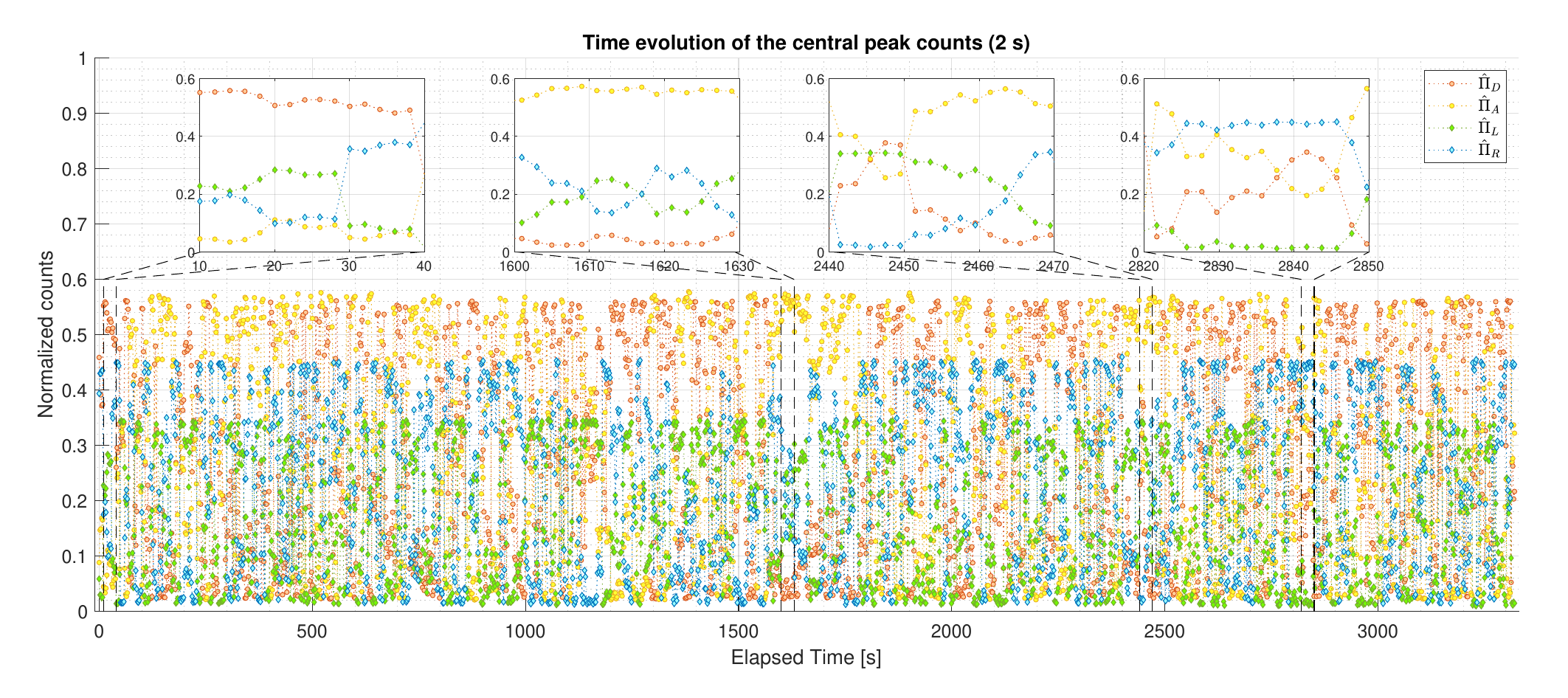}
    \caption{Collected counts during 1 hour of acquisition (main plot) together with four detailed frames highlighting the different correlations along the data-take. Each point in the graph is derived considering 2 s of integration window.}
    \label{fig:collectedCounts_2s}
\end{figure*}

\begin{table}[]
    \centering
    \setlength{\tabcolsep}{10pt} 
    \renewcommand{\arraystretch}{1.5} 
    \begin{tabular}{|c|c|}
        \hline
        \textbf{Elements} & \textbf{Fidelity} \\
        \hline
        \hline
        $\mathcal{F}(\Pi_D^\mathrm{id},\Pi_D^\mathrm{exp})$ & 0.971 \\
        \hline
        $\mathcal{F}(\Pi_A^\mathrm{id},\Pi_A^\mathrm{exp})$ & 0.942 \\
        \hline
        $\mathcal{F}(\Pi_L^\mathrm{id},\Pi_L^\mathrm{exp})$ & 0.979 \\
        \hline
        $\mathcal{F}(\Pi_R^\mathrm{id},\Pi_R^\mathrm{exp})$ & 0.948 \\
        \hline
    \end{tabular}
    \caption{Fidelity between the ideal  and experimental  POVMs computed according to \cref{eq:fidelity}.}
    \label{tab:fidelity}
\end{table}

\subsection{Proof-of-Principle RFI-QKD experiment}
A data acquisition of 1 hour was performed for the Proof-of-Principle RFI-QKD experiment.\\
The collected counts are pictured in \cref{fig:collectedCounts_2s}, where one can see the evolution in the entire experiment together with four details of $30$ seconds of the experimental run, which can better clarify the good level of anti-correlation between the states belonging to the same basis and the mutual uncorrelation between the selected measurement bases. Measurement apparatus defects are mainly noticeable in the maximum excursion of each channel, caused by the different loss factors for each output branch. In particular, the beam splitter used to randomly choose between measurements in the $\mathbb{X}$ or $\mathbb{Y}$ bases exhibited a 55:45 split ratio, and the $\Pi_L$ detector had a 30\% lower coupling and detection efficiency compared to the others. However, this condition is not detrimental for the final results, but on the contrary proves the feasibility of this strategy with realistic non-ideal devices. Regarding the choice of the integration time interval, we tested different \textit{integration windows} and the one that gave the slightly best result was of $2$ seconds when considering finite-key effects.

\begin{figure}
    \centering
    \includegraphics[width=0.99\linewidth]{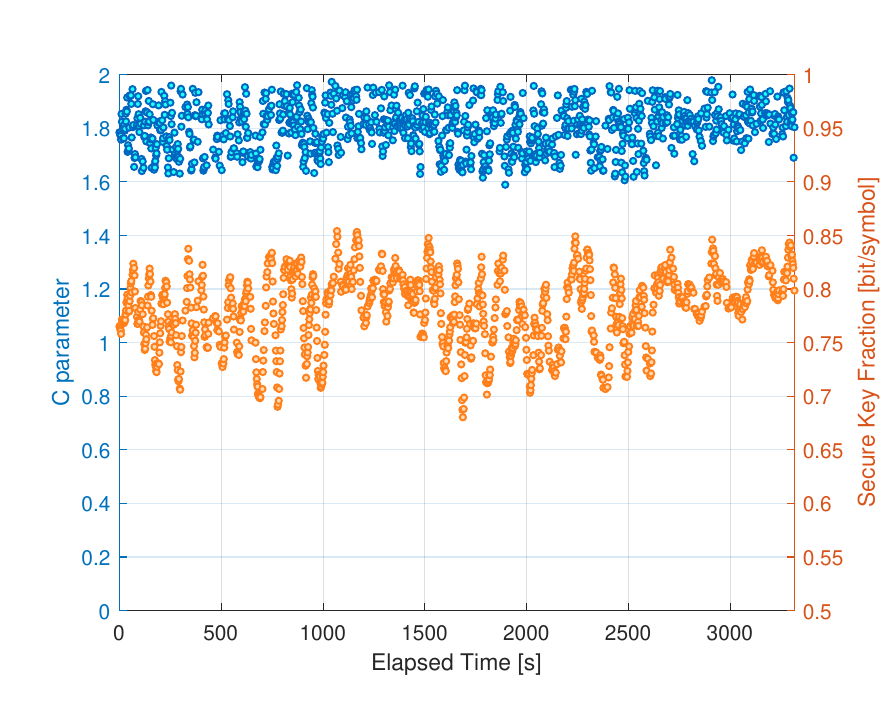}
    \caption{Evolution of the C parameter and the secret key fraction obtained in the proof-of-principle RFI-QKD experiment.}
    \label{fig:c_skr_2s}
\end{figure}

The security parameter C and the secret key fraction were calculated following the procedure described in \cref{sec:security}. The results are reported in \cref{fig:c_skr_2s}. We obtain an average C value of $1.80 \pm 0.09$. This leads on average to a maximum of $0.78 \pm 0.06$ secure bits per exchanged key symbol between transmitter and receiver. These outcome validate our approach to RFI-QKD, while exhibiting results that are compatible in terms of performance with other state-of-the-art demonstrations~\cite{RFI_QKD_china,Wang2019, Chen2020, Tang2022}.

\section{Conclusions}
In summary, this work introduces a novel, fully-passive and self-characterizing cross-encoded receiver for Reference-Frame-Independent Quantum Key Distribution (RFI-QKD). By leveraging a time-bin to polarization conversion technique and employing a Quantum Detector Self-Characterization (QDSC) method, we have demonstrated a substantial reduction in implementation complexity and enhanced security in quantum cryptographic communications. Our proposed receiver eliminates the need for active phase modulation and real-time synchronization, which traditionally complicate QKD systems, while maintaining robustness against polarization mode dispersion and other channel imperfections.

The integration of QDSC into the security analysis of the QKD protocol represents a significant advancement, as it allows for a realistic and comprehensive characterization of the measurement apparatus, incorporating actual physical defects and variations. This ensures a higher level of secrecy and reliability, moving beyond idealized device models commonly used in previous studies.

The experimental implementation of our receiver confirms its feasibility and potential for real-world applications, highlighting its capability to obtain high values for the RFI-QKD security parameter $C$. This advancement paves the way for more practical and secure quantum communication systems, contributing to the broader adoption of quantum cryptography in various critical sectors.

The implementation of field trails in relevant environments is the natural evolution of this research work. These experiments would further validate our approach to time-bin encoded RFI-QKD and would interface our receiver with active QKD transmitters. Future research could also explore further optimizations of the passive receiver design and the extension of the QDSC technique to other quantum cryptographic protocols, enhancing their security and practicality. In fact, the continued development and integration of innovative techniques will be crucial in addressing the evolving challenges in the field of quantum communication and cybersecurity.

\section*{Acknowledgments}
 \noindent M.G.'s PhD scholarship was co-funded by \textit{Telebit S.p.A.} and MIUR - Ministerial Decree 352/2022. This work was partially funded by European Union’s Horizon Europe research and innovation programme under the project Quantum Secure Networks Partnership (QSNP, grant agreement No 101114043), by Ministero della Difesa (Q4SEC, Contratto N. 729 di Rep. del 09.03.2023 - CIG: Z692F967FF) and by Agenzia Spaziale Italiana (2018-14-HH.0, CUP: E16J16001490001, Q-SecGroundSpace).\\ Views and opinions expressed are, however, those of the author(s) only and do not necessarily reflect those of the European Union or the European Commission-EU. Neither the European Union nor the granting authority can be held responsible for them.

\appendix

\section{Ellipsoid fit}\label{sec:appendixA}
\subsection{Derivation of Ellipsoid parameters from POVM elements}
Consider a density operator $\rho$ and the elements $\{\pi_k\}$ of a POVM  which can be written in the Bloch sphere notation respectively as
\begin{equation}
    \rho = \frac{1}{2}(\mathds{1} + \textbf{r}\cdot\boldsymbol{\sigma})
\end{equation}

\begin{equation}\label{eq:POVM_system}
    \pi_k = t_k\mathds{1} + \textbf{m}_k\cdot\boldsymbol{\sigma}
\end{equation}
where $\boldsymbol{\sigma} = (\sigma_x,\sigma_y,\sigma_z)$ is the Pauli tensor of Pauli matrices. The vector $\textbf{r} = (r_x, r_y, r_z)$ is the Bloch vector related to $\rho$, having the physical requirement to satisfy the positivity constraint $|\textbf{r}|^2 \leq 1$. Instead, the operators $\pi_k$ relate to the vector $t_k$ and $\textbf{m}_k = (m_{k,x}, m_{k,y}, m_{k,z})$ that 
must satisfy the unitarity constraint on the POVMs $\sum_k \pi_k = \mathds{1}$, which implies that $m_{k,x}+m_{k,y}+m_{k,z} = 0$ and $\sum_k t_k = 1$ while $t_k > |\textbf{m}_k|$ from $\pi_k \geq 0$. Therefore, according to Born's rule, a general constraint can be introduced:
\begin{equation}
    p_k = Tr(\rho \pi_k) = t_k + \textbf{m}_k\cdot\textbf{r}
\end{equation}
that can be written in matrix form as
\begin{equation}
    (\textbf{p} - \textbf{t}) = M_{n \times 3}\cdot \textbf{r}.
\end{equation}
In this shape, the positivity constraint about \textbf{r} can be rearranged into a constraint on \textbf{p}, and therefore
\begin{equation}
    1 \geq |\textbf{r}|^2 = \textbf{r}^T\textbf{r} = (\textbf{p} - \textbf{t})^T (M^+)^T M^+ (\textbf{p} - \textbf{t})
\end{equation}
which, after the definition $Q = M\cdot M^T$, can be rewritten as in \cref{eq:POVMsDecomposition}
\begin{equation*}
    (\textbf{p} - \textbf{t})^T Q^+ (\textbf{p} - \textbf{t}) \leq 1.
\end{equation*}
which describes an hyper-ellipsoid lying in an \textbf{n}-dimensional space, determined by the matrix $Q^+$ centered in \textbf{t}.

\subsection{Dimensional reduction of the problem}

As a first step, the Principle Component Analysis requires the removal of the  average probability over the different \textit{m} probe states, thus obtaining a new matrix \textit{A}, described as
\begin{equation}
    A_{n \times m} = \begin{pmatrix}
        p_0^{(0)}-\bar{p}_0 & \dots & p_0^{(m-1)}-\bar{p}_0 \\
        p_1^{(0)}-\bar{p}_1 &  \dots & p_1^{(m-1)}-\bar{p}_1 \\
        \vdots  & \ddots & \vdots \\
        p_{n-1}^{(0)}-\bar{p}_{n-1} & \dots & p_{n-1}^{(m-1)}-\bar{p}_{n-1} \\
    \end{pmatrix},
\end{equation}
where each mean value has the form
\begin{equation}
    \bar{p}_k = \frac{1}{m}\sum_{j=0}^{m-1}p_k^{(j)}.
\end{equation}
The next step is to remove redundant linear dependent outcomes and extract valuable features, by means of singular value decomposition (SVD). A standard SVD can be generally described in the following way:
\begin{equation}\label{eq:complete_svd}
    A_{n \times m} = U \cdot \Sigma \cdot V^T = U_{n \times n} \cdot
    \begin{pmatrix}
        s_1 & & \\
        & \ddots & \\
        & & s_n
    \end{pmatrix}_{n \times m}
    \cdot V_{m \times m}^T.
\end{equation}
Since the response range $\mathcal{W}$ should lie in an affine plane up to dimensionality \textbf{3} thanks to the linear dependencies of the measurement operators, the final SVD can be simplified considering a \textbf{3}-dimensional $\Sigma$  matrix, with grater order elements of order $O(1/\sqrt{N})$, where $N$ represents the sum of the collected counts of measurement outcome for each probe state. The problem thus simplifies as
\begin{equation}
    A_{n \times m} = U_{n \times 3} \cdot
    \begin{pmatrix}
        s_1 & & \\
        & s_2 & \\
        & & s_3
    \end{pmatrix}
    \cdot V_{3 \times m}^T
\end{equation}
with the reduced matrix $\Tilde{A}$ derivable according to the reverse equation
\begin{equation}\label{eq:reduced_svd}
     \Tilde{A}_{3 \times m} = (U^T)_{3 \times n} \cdot A_{n \times m} =
    \begin{pmatrix}
        s_1 & & \\
        & s_2 & \\
        & & s_3
    \end{pmatrix}
    \cdot V_{3 \times m}^T.
\end{equation}
The passage described from \cref{eq:complete_svd} to \cref{eq:reduced_svd} is crucial to reduce the dimensionality of the data set and thus the complexity of the considered problem. This is compatible with the principal component analysis that can be implemented with SVD.\\ The retrieved matrix can be seen as composed by \textbf{3}-dimensional column vectors, each one representing a point in the spatial domain of the matrix $\Tilde{A}_{3 \times m}$. Given that the response range $\mathcal{W}(\pi)$ is a convex set, and therefore each inner point can be obtained by means of linear combination of the external boundary coordinates, we are only interested in the boundary data of the whole set in order to describe it. The boundary data set of this reduced 3-dimensional problem has the following shape
\begin{equation}
    \begin{split}
    \mathcal{B} &= \Bigl\{ \textbf{v}_j = \Bigl(\Tilde{A}(1,j), \Tilde{A}(2,j), \Tilde{A}(3,j)\Bigr)^T, \\ &\text{with $\textbf{v}_j \in$ CHB}\Bigl( \{\textbf{v}_j | j=0, 1, \dots, m-1\} \Bigr) \Bigr\}
\end{split}
\end{equation}
where \textit{CHB} stands for \textbf{Convex-Hull Boundary}.\\ The new boundary data set can thus be described as the reduced set $B_{n \times m'}$.

The last step is to reconstruct an ellipsoid compatible with the space $\Tilde{p} = U \cdot (p-\bar{p})$ by means of a direct ellipsoid fitting equation, built on the points collected in $B_{n \times m'}$.

Since in our case the number of linear independent elements of a qubit POVM is \textbf{2}, the retrieved response range set is translated into an ellipse lying on a \textbf{2}-dimensional plane. It is important to stress the fact that $Q$ and \textbf{t} determine a representation of the structure of the POVM elements which fully characterizes the physical model of the receiver apparatus, including the experimental errors introduced by the measurement system.

To efficiently derive the experimental POVMs it is necessary to have a precise geometrical description of the ellipse associated with the probability space of the collected measurements. Therefore we performed an ellipsoid fit over the reduced boundary data set $B_{n \times m'}$ (see \cref{fig:ellipseFit}), according to the generic ellipsoid equation once the third variable ($z$ in this case) has been set to zero:
\begin{equation}
    \begin{cases}
      \begin{split}
          \quad ax^2 + by^2 + cz^2 + dxy + eyz +\quad\quad  \\ + fxz + gx + hy + iz + j = 0
      \end{split}   \\
        \quad z = 0.
    \end{cases}
\end{equation}
The matrix equation of a generic ellipsoid centered in \textbf{w} is given by
\begin{equation}
    (\textbf{v}-\textbf{w})^TA(\textbf{v}-\textbf{w}) = C
\end{equation}
considering the followings:
\begin{equation}
    \begin{aligned}
        \textbf{v} &= \begin{pmatrix} x\\ y\\ z \end{pmatrix},\quad 
        A = \begin{pmatrix} a & \frac{d}{2} & \frac{f}{2} \\ \frac{d}{2} &  b & \frac{e}{2} \\ \frac{f}{2} & \frac{e}{2} &  c \end{pmatrix}, \\
        \textbf{w} &= A^{-1} \Bigl(-\frac{1}{2}\Bigr) \begin{pmatrix} g \\ h \\ i \end{pmatrix},\quad 
        C = \textbf{w}^TA\textbf{w} - j.
    \end{aligned}
\end{equation}
Considering the association $A \rightarrow Q_{3\times3}$ and $\bar{\textbf{p}} \rightarrow \textbf{t}$, we gain a direct relation between the geometric problem and the POVMs derivation problem. 

Given a direct relation between the geometric elliptical description and the POVMs decomposition shown in \cref{eq:POVMsDecomposition} and \cref{eq:POVMsDescription}, it is possible to reconstruct the ellipsoid in the $\Tilde{p}_{4 \times 4}$ space simply applying a space transformation, considering $U_{n \times 3}$ matrix from \cref{eq:reduced_svd} as follows
\begin{equation}\label{eq:4dim_map}
    Q_{4 \times 4} = - \biggl( U_{n \times 3} \cdot Q_{3 \times 3} \cdot (U^+)_{3 \times n} \biggr)^+ \Bigr|_{n=4}
\end{equation}

Finally, we have been able to retrieve the shape of the sought POVMs by solving the linear system described in \cref{eq:POVM_system}, now having $Q$ and \textbf{t} as the known parameters.

\section{Further detail on the Experimental POVM Determination}\label{sec:appendixB}

\begin{figure}
     \centering
     \begin{subfigure}{0.49\textwidth}
         \centering
         \includegraphics[width=0.75\linewidth]{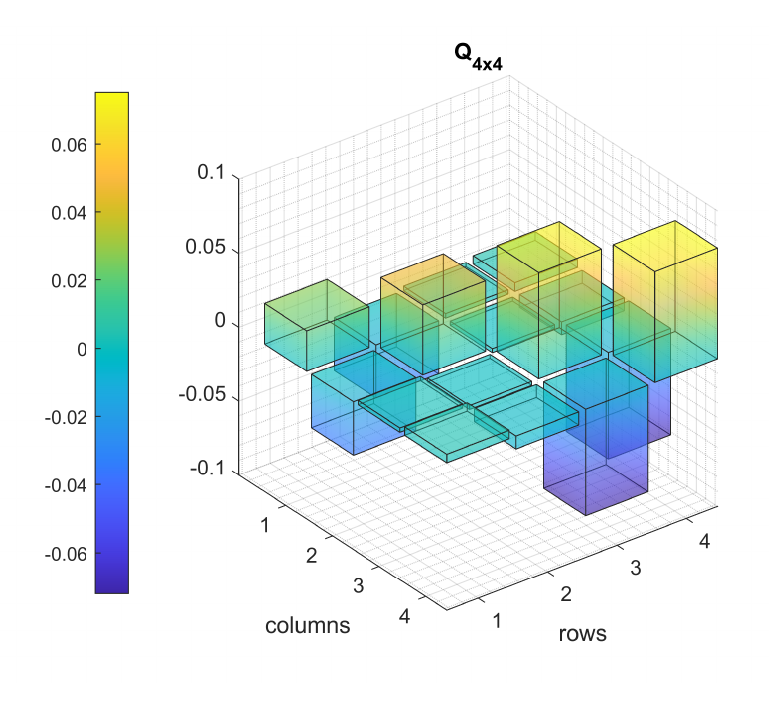}
     \end{subfigure}
     \hfill
     \begin{subfigure}{0.49\textwidth}
        \centering
        \includegraphics[width=0.75\linewidth]{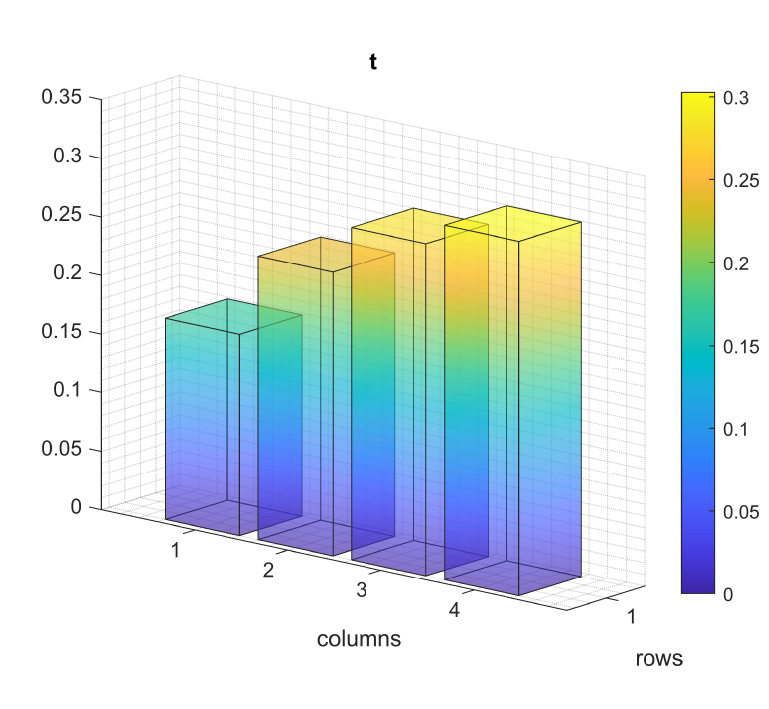}
     \end{subfigure}
     \caption{Bar representation of the $Q_{4 \times 4}$ matrix (\textit{left}) and the \textbf{t} vector (\textit{right}) computed with the explained process.}
     \label{fig:Qt_bars}
\end{figure}

The $Q_{3 \times 3}$ matrix obtained after the ellipse fit in the \textbf{2}-dim space has the following shape:
\begin{equation*}
    Q_{3 \times 3} = \begin{pmatrix}
        -6.9176 & 0.1565 & 0 \\ 0.1565 & -13.2780 & 0 \\ 0 & 0 & 0
    \end{pmatrix}.
\end{equation*}
After the application of \cref{eq:4dim_map} we derived the following quantities:
\begin{align*}
     Q_{4 \times 4} &= \begin{pmatrix}
        0.0272 & -0.0358 & 0.0029 & 0.0057 \\
        -0.0358 & 0.0471 & -0.0026 & -0.0087 \\
        0.0029 & -0.0026 & 0.0712 & -0.0715 \\
        0.0057 & -0.0087 & -0.0715 & 0.0744
    \end{pmatrix}, \\
    \textbf{t} &= \begin{pmatrix}
        0.1717 \\ 0.2420 \\ 0.2836 \\ 0.3027
    \end{pmatrix}.
\end{align*}
A figurative representation of these two quantities is reported for clarity in \cref{fig:Qt_bars}.

According to the theory, the representation of the sought POVMs can be defined up to a reference frame specification. In principle, this passage is not mandatory for the realization of the protocol, but reduces the complexity of the equations system to be solved. In our scenario, we chose the physical receiver implementing the action of the measurement operators on the collected photons to equal a particular reference frame made of the standard basis in Pauli notation $\boldsymbol{\sigma}_x$. We also made the $\hat{x}$ direction of the reference frame to be parallel to the $\boldsymbol{m}_2$ vector. In the computation of the system discussed in \cref{eq:POVM_system}, this translated in the following definitions:
\begin{equation*}
    \begin{cases}
        m_{0,x} = 0 \\ m_{2,y} = m_{2,z} = 0
    \end{cases}
\end{equation*}

From this, the POVMs associated with the exploited receiver (depicted in \cref{fig:povm_bars}), are obtained:
\begin{align*}
    \hat{\Pi}_L &= \begin{pmatrix} +0.1718 + 0.0000i & +0.0106 - 0.1645i \\ +0.0106 + 0.1645i & +0.1718 + 0.0000i \end{pmatrix}\\
    \hat{\Pi}_R &= \begin{pmatrix} +0.2465 + 0.0000i & -0.0099 + 0.2168i \\ -0.0099 - 0.2168i & +0.2375 + 0.0000i \end{pmatrix}\\
    \hat{\Pi}_D &= \begin{pmatrix} +0.2836 + 0.0000i &  +0.2669 + 0.0000i \\ +0.2669 + 0.0000i & +0.2836 + 0.0000i \end{pmatrix}\\
    \hat{\Pi}_A &= \begin{pmatrix} +0.2923 + 0.0000i & -0.2676 - 0.0522i \\ -0.2676 + 0.0522i & +0.2973 + 0.0000i \end{pmatrix}.
\end{align*}

\bibliography{bibliography}

\providecommand{\noopsort}[1]{}\providecommand{\singleletter}[1]{#1}%
\begin{thebibliography}{40}%
\makeatletter
\providecommand \@ifxundefined [1]{%
 \@ifx{#1\undefined}
}%
\providecommand \@ifnum [1]{%
 \ifnum #1\expandafter \@firstoftwo
 \else \expandafter \@secondoftwo
 \fi
}%
\providecommand \@ifx [1]{%
 \ifx #1\expandafter \@firstoftwo
 \else \expandafter \@secondoftwo
 \fi
}%
\providecommand \natexlab [1]{#1}%
\providecommand \enquote  [1]{``#1''}%
\providecommand \bibnamefont  [1]{#1}%
\providecommand \bibfnamefont [1]{#1}%
\providecommand \citenamefont [1]{#1}%
\providecommand \href@noop [0]{\@secondoftwo}%
\providecommand \href [0]{\begingroup \@sanitize@url \@href}%
\providecommand \@href[1]{\@@startlink{#1}\@@href}%
\providecommand \@@href[1]{\endgroup#1\@@endlink}%
\providecommand \@sanitize@url [0]{\catcode `\\12\catcode `\$12\catcode `\&12\catcode `\#12\catcode `\^12\catcode `\_12\catcode `\%12\relax}%
\providecommand \@@startlink[1]{}%
\providecommand \@@endlink[0]{}%
\providecommand \url  [0]{\begingroup\@sanitize@url \@url }%
\providecommand \@url [1]{\endgroup\@href {#1}{\urlprefix }}%
\providecommand \urlprefix  [0]{URL }%
\providecommand \Eprint [0]{\href }%
\providecommand \doibase [0]{https://doi.org/}%
\providecommand \selectlanguage [0]{\@gobble}%
\providecommand \bibinfo  [0]{\@secondoftwo}%
\providecommand \bibfield  [0]{\@secondoftwo}%
\providecommand \translation [1]{[#1]}%
\providecommand \BibitemOpen [0]{}%
\providecommand \bibitemStop [0]{}%
\providecommand \bibitemNoStop [0]{.\EOS\space}%
\providecommand \EOS [0]{\spacefactor3000\relax}%
\providecommand \BibitemShut  [1]{\csname bibitem#1\endcsname}%
\let\auto@bib@innerbib\@empty
\bibitem [{\citenamefont {Bennett}\ and\ \citenamefont {Brassard}(2014)}]{BB84}%
  \BibitemOpen
  \bibfield  {author} {\bibinfo {author} {\bibfnamefont {C.~H.}\ \bibnamefont {Bennett}}\ and\ \bibinfo {author} {\bibfnamefont {G.}~\bibnamefont {Brassard}},\ }\bibfield  {title} {\bibinfo {title} {{Quantum cryptography: Public key distribution and coin tossing}},\ }\href {https://doi.org/10.1016/j.tcs.2014.05.025} {\bibfield  {journal} {\bibinfo  {journal} {Theor. Comput. Sci.}\ }\textbf {\bibinfo {volume} {560}},\ \bibinfo {pages} {7} (\bibinfo {year} {2014})}\BibitemShut {NoStop}%
\bibitem [{\citenamefont {Shor}(1997)}]{Shor1997}%
  \BibitemOpen
  \bibfield  {author} {\bibinfo {author} {\bibfnamefont {P.}~\bibnamefont {Shor}},\ }\bibfield  {title} {\bibinfo {title} {Polynomial-time algorithms for prime factorization and discrete logarithms on a quantum computer},\ }\href {https://doi.org/10.1137/S0097539795293172} {\bibfield  {journal} {\bibinfo  {journal} {SIAM J. Comput.}\ }\textbf {\bibinfo {volume} {26}},\ \bibinfo {pages} {1484} (\bibinfo {year} {1997})}\BibitemShut {NoStop}%
\bibitem [{\citenamefont {Bouland}\ \emph {et~al.}(2019)\citenamefont {Bouland}, \citenamefont {Fefferman}, \citenamefont {Nirkhe},\ and\ \citenamefont {Vazirani}}]{QC_Bouland2019}%
  \BibitemOpen
  \bibfield  {author} {\bibinfo {author} {\bibfnamefont {A.}~\bibnamefont {Bouland}}, \bibinfo {author} {\bibfnamefont {B.}~\bibnamefont {Fefferman}}, \bibinfo {author} {\bibfnamefont {C.}~\bibnamefont {Nirkhe}},\ and\ \bibinfo {author} {\bibfnamefont {U.}~\bibnamefont {Vazirani}},\ }\bibfield  {title} {\bibinfo {title} {On the complexity and verification of quantum random circuit sampling},\ }\href {https://doi.org/10.1038/s41567-018-0318-2} {\bibfield  {journal} {\bibinfo  {journal} {Nature Physics}\ }\textbf {\bibinfo {volume} {15}},\ \bibinfo {pages} {159} (\bibinfo {year} {2019})}\BibitemShut {NoStop}%
\bibitem [{\citenamefont {Arute}\ \emph {et~al.}(2019)\citenamefont {Arute}, \citenamefont {Arya}, \citenamefont {Babbush}, \citenamefont {Bacon}, \citenamefont {Bardin}, \citenamefont {Barends}, \citenamefont {Biswas}, \citenamefont {Boixo}, \citenamefont {Brandao}, \citenamefont {Buell}, \citenamefont {Burkett}, \citenamefont {Chen}, \citenamefont {Chen}, \citenamefont {Chiaro}, \citenamefont {Collins}, \citenamefont {Courtney}, \citenamefont {Dunsworth}, \citenamefont {Farhi}, \citenamefont {Foxen}, \citenamefont {Fowler}, \citenamefont {Gidney}, \citenamefont {Giustina}, \citenamefont {Graff}, \citenamefont {Guerin}, \citenamefont {Habegger}, \citenamefont {Harrigan}, \citenamefont {Hartmann}, \citenamefont {Ho}, \citenamefont {Hoffmann}, \citenamefont {Huang}, \citenamefont {Humble}, \citenamefont {Isakov}, \citenamefont {Jeffrey}, \citenamefont {Jiang}, \citenamefont {Kafri}, \citenamefont {Kechedzhi}, \citenamefont {Kelly}, \citenamefont {Klimov}, \citenamefont {Knysh}, \citenamefont {Korotkov},
  \citenamefont {Kostritsa}, \citenamefont {Landhuis}, \citenamefont {Lindmark}, \citenamefont {Lucero}, \citenamefont {Lyakh}, \citenamefont {Mandrà}, \citenamefont {McClean}, \citenamefont {McEwen}, \citenamefont {Megrant}, \citenamefont {Mi}, \citenamefont {Michielsen}, \citenamefont {Mohseni}, \citenamefont {Mutus}, \citenamefont {Naaman}, \citenamefont {Neeley}, \citenamefont {Neill}, \citenamefont {Niu}, \citenamefont {Ostby}, \citenamefont {Petukhov}, \citenamefont {Platt}, \citenamefont {Quintana}, \citenamefont {Rieffel}, \citenamefont {Roushan}, \citenamefont {Rubin}, \citenamefont {Sank}, \citenamefont {Satzinger}, \citenamefont {Smelyanskiy}, \citenamefont {Sung}, \citenamefont {Trevithick}, \citenamefont {Vainsencher}, \citenamefont {Villalonga}, \citenamefont {White}, \citenamefont {Yao}, \citenamefont {Yeh}, \citenamefont {Zalcman}, \citenamefont {Neven},\ and\ \citenamefont {Martinis}}]{GoogleQC2019}%
  \BibitemOpen
  \bibfield  {author} {\bibinfo {author} {\bibfnamefont {F.}~\bibnamefont {Arute}}, \bibinfo {author} {\bibfnamefont {K.}~\bibnamefont {Arya}}, \bibinfo {author} {\bibfnamefont {R.}~\bibnamefont {Babbush}}, \bibinfo {author} {\bibfnamefont {D.}~\bibnamefont {Bacon}}, \bibinfo {author} {\bibfnamefont {J.~C.}\ \bibnamefont {Bardin}}, \bibinfo {author} {\bibfnamefont {R.}~\bibnamefont {Barends}}, \bibinfo {author} {\bibfnamefont {R.}~\bibnamefont {Biswas}}, \bibinfo {author} {\bibfnamefont {S.}~\bibnamefont {Boixo}}, \bibinfo {author} {\bibfnamefont {F.~G. S.~L.}\ \bibnamefont {Brandao}}, \bibinfo {author} {\bibfnamefont {D.~A.}\ \bibnamefont {Buell}}, \bibinfo {author} {\bibfnamefont {B.}~\bibnamefont {Burkett}}, \bibinfo {author} {\bibfnamefont {Y.}~\bibnamefont {Chen}}, \bibinfo {author} {\bibfnamefont {Z.}~\bibnamefont {Chen}}, \bibinfo {author} {\bibfnamefont {B.}~\bibnamefont {Chiaro}}, \bibinfo {author} {\bibfnamefont {R.}~\bibnamefont {Collins}}, \bibinfo {author} {\bibfnamefont {W.}~\bibnamefont
  {Courtney}}, \bibinfo {author} {\bibfnamefont {A.}~\bibnamefont {Dunsworth}}, \bibinfo {author} {\bibfnamefont {E.}~\bibnamefont {Farhi}}, \bibinfo {author} {\bibfnamefont {B.}~\bibnamefont {Foxen}}, \bibinfo {author} {\bibfnamefont {A.}~\bibnamefont {Fowler}}, \bibinfo {author} {\bibfnamefont {C.}~\bibnamefont {Gidney}}, \bibinfo {author} {\bibfnamefont {M.}~\bibnamefont {Giustina}}, \bibinfo {author} {\bibfnamefont {R.}~\bibnamefont {Graff}}, \bibinfo {author} {\bibfnamefont {K.}~\bibnamefont {Guerin}}, \bibinfo {author} {\bibfnamefont {S.}~\bibnamefont {Habegger}}, \bibinfo {author} {\bibfnamefont {M.~P.}\ \bibnamefont {Harrigan}}, \bibinfo {author} {\bibfnamefont {M.~J.}\ \bibnamefont {Hartmann}}, \bibinfo {author} {\bibfnamefont {A.}~\bibnamefont {Ho}}, \bibinfo {author} {\bibfnamefont {M.}~\bibnamefont {Hoffmann}}, \bibinfo {author} {\bibfnamefont {T.}~\bibnamefont {Huang}}, \bibinfo {author} {\bibfnamefont {T.~S.}\ \bibnamefont {Humble}}, \bibinfo {author} {\bibfnamefont {S.~V.}\ \bibnamefont
  {Isakov}}, \bibinfo {author} {\bibfnamefont {E.}~\bibnamefont {Jeffrey}}, \bibinfo {author} {\bibfnamefont {Z.}~\bibnamefont {Jiang}}, \bibinfo {author} {\bibfnamefont {D.}~\bibnamefont {Kafri}}, \bibinfo {author} {\bibfnamefont {K.}~\bibnamefont {Kechedzhi}}, \bibinfo {author} {\bibfnamefont {J.}~\bibnamefont {Kelly}}, \bibinfo {author} {\bibfnamefont {P.~V.}\ \bibnamefont {Klimov}}, \bibinfo {author} {\bibfnamefont {S.}~\bibnamefont {Knysh}}, \bibinfo {author} {\bibfnamefont {A.}~\bibnamefont {Korotkov}}, \bibinfo {author} {\bibfnamefont {F.}~\bibnamefont {Kostritsa}}, \bibinfo {author} {\bibfnamefont {D.}~\bibnamefont {Landhuis}}, \bibinfo {author} {\bibfnamefont {M.}~\bibnamefont {Lindmark}}, \bibinfo {author} {\bibfnamefont {E.}~\bibnamefont {Lucero}}, \bibinfo {author} {\bibfnamefont {D.}~\bibnamefont {Lyakh}}, \bibinfo {author} {\bibfnamefont {S.}~\bibnamefont {Mandrà}}, \bibinfo {author} {\bibfnamefont {J.~R.}\ \bibnamefont {McClean}}, \bibinfo {author} {\bibfnamefont {M.}~\bibnamefont {McEwen}},
  \bibinfo {author} {\bibfnamefont {A.}~\bibnamefont {Megrant}}, \bibinfo {author} {\bibfnamefont {X.}~\bibnamefont {Mi}}, \bibinfo {author} {\bibfnamefont {K.}~\bibnamefont {Michielsen}}, \bibinfo {author} {\bibfnamefont {M.}~\bibnamefont {Mohseni}}, \bibinfo {author} {\bibfnamefont {J.}~\bibnamefont {Mutus}}, \bibinfo {author} {\bibfnamefont {O.}~\bibnamefont {Naaman}}, \bibinfo {author} {\bibfnamefont {M.}~\bibnamefont {Neeley}}, \bibinfo {author} {\bibfnamefont {C.}~\bibnamefont {Neill}}, \bibinfo {author} {\bibfnamefont {M.~Y.}\ \bibnamefont {Niu}}, \bibinfo {author} {\bibfnamefont {E.}~\bibnamefont {Ostby}}, \bibinfo {author} {\bibfnamefont {A.}~\bibnamefont {Petukhov}}, \bibinfo {author} {\bibfnamefont {J.~C.}\ \bibnamefont {Platt}}, \bibinfo {author} {\bibfnamefont {C.}~\bibnamefont {Quintana}}, \bibinfo {author} {\bibfnamefont {E.~G.}\ \bibnamefont {Rieffel}}, \bibinfo {author} {\bibfnamefont {P.}~\bibnamefont {Roushan}}, \bibinfo {author} {\bibfnamefont {N.~C.}\ \bibnamefont {Rubin}}, \bibinfo
  {author} {\bibfnamefont {D.}~\bibnamefont {Sank}}, \bibinfo {author} {\bibfnamefont {K.~J.}\ \bibnamefont {Satzinger}}, \bibinfo {author} {\bibfnamefont {V.}~\bibnamefont {Smelyanskiy}}, \bibinfo {author} {\bibfnamefont {K.~J.}\ \bibnamefont {Sung}}, \bibinfo {author} {\bibfnamefont {M.~D.}\ \bibnamefont {Trevithick}}, \bibinfo {author} {\bibfnamefont {A.}~\bibnamefont {Vainsencher}}, \bibinfo {author} {\bibfnamefont {B.}~\bibnamefont {Villalonga}}, \bibinfo {author} {\bibfnamefont {T.}~\bibnamefont {White}}, \bibinfo {author} {\bibfnamefont {Z.~J.}\ \bibnamefont {Yao}}, \bibinfo {author} {\bibfnamefont {P.}~\bibnamefont {Yeh}}, \bibinfo {author} {\bibfnamefont {A.}~\bibnamefont {Zalcman}}, \bibinfo {author} {\bibfnamefont {H.}~\bibnamefont {Neven}},\ and\ \bibinfo {author} {\bibfnamefont {J.~M.}\ \bibnamefont {Martinis}},\ }\bibfield  {title} {\bibinfo {title} {Quantum supremacy using a programmable superconducting processor},\ }\href {https://doi.org/10.1038/s41586-019-1666-5} {\bibfield  {journal}
  {\bibinfo  {journal} {Nature}\ }\textbf {\bibinfo {volume} {574}},\ \bibinfo {pages} {505} (\bibinfo {year} {2019})}\BibitemShut {NoStop}%
\bibitem [{\citenamefont {Scarani}\ \emph {et~al.}(2009)\citenamefont {Scarani}, \citenamefont {Bechmann-Pasquinucci}, \citenamefont {Cerf}, \citenamefont {Du\ifmmode~\check{s}\else \v{s}\fi{}ek}, \citenamefont {L\"utkenhaus},\ and\ \citenamefont {Peev}}]{Scarani2008}%
  \BibitemOpen
  \bibfield  {author} {\bibinfo {author} {\bibfnamefont {V.}~\bibnamefont {Scarani}}, \bibinfo {author} {\bibfnamefont {H.}~\bibnamefont {Bechmann-Pasquinucci}}, \bibinfo {author} {\bibfnamefont {N.~J.}\ \bibnamefont {Cerf}}, \bibinfo {author} {\bibfnamefont {M.}~\bibnamefont {Du\ifmmode~\check{s}\else \v{s}\fi{}ek}}, \bibinfo {author} {\bibfnamefont {N.}~\bibnamefont {L\"utkenhaus}},\ and\ \bibinfo {author} {\bibfnamefont {M.}~\bibnamefont {Peev}},\ }\bibfield  {title} {\bibinfo {title} {The security of practical quantum key distribution},\ }\href {https://doi.org/10.1103/RevModPhys.81.1301} {\bibfield  {journal} {\bibinfo  {journal} {Rev. Mod. Phys.}\ }\textbf {\bibinfo {volume} {81}},\ \bibinfo {pages} {1301} (\bibinfo {year} {2009})}\BibitemShut {NoStop}%
\bibitem [{\citenamefont {Pirandola}\ \emph {et~al.}(2020)\citenamefont {Pirandola}, \citenamefont {Andersen}, \citenamefont {Banchi}, \citenamefont {Berta}, \citenamefont {Bunandar}, \citenamefont {Colbeck}, \citenamefont {Englund}, \citenamefont {Gehring}, \citenamefont {Lupo}, \citenamefont {Ottaviani}, \citenamefont {Pereira}, \citenamefont {Razavi}, \citenamefont {{Shamsul Shaari}}, \citenamefont {Tomamichel}, \citenamefont {Usenko}, \citenamefont {Vallone}, \citenamefont {Villoresi},\ and\ \citenamefont {Wallden}}]{Pirandola2019rev}%
  \BibitemOpen
  \bibfield  {author} {\bibinfo {author} {\bibfnamefont {S.}~\bibnamefont {Pirandola}}, \bibinfo {author} {\bibfnamefont {U.~L.}\ \bibnamefont {Andersen}}, \bibinfo {author} {\bibfnamefont {L.}~\bibnamefont {Banchi}}, \bibinfo {author} {\bibfnamefont {M.}~\bibnamefont {Berta}}, \bibinfo {author} {\bibfnamefont {D.}~\bibnamefont {Bunandar}}, \bibinfo {author} {\bibfnamefont {R.}~\bibnamefont {Colbeck}}, \bibinfo {author} {\bibfnamefont {D.}~\bibnamefont {Englund}}, \bibinfo {author} {\bibfnamefont {T.}~\bibnamefont {Gehring}}, \bibinfo {author} {\bibfnamefont {C.}~\bibnamefont {Lupo}}, \bibinfo {author} {\bibfnamefont {C.}~\bibnamefont {Ottaviani}}, \bibinfo {author} {\bibfnamefont {J.~L.}\ \bibnamefont {Pereira}}, \bibinfo {author} {\bibfnamefont {M.}~\bibnamefont {Razavi}}, \bibinfo {author} {\bibfnamefont {J.}~\bibnamefont {{Shamsul Shaari}}}, \bibinfo {author} {\bibfnamefont {M.}~\bibnamefont {Tomamichel}}, \bibinfo {author} {\bibfnamefont {V.~C.}\ \bibnamefont {Usenko}}, \bibinfo {author} {\bibfnamefont
  {G.}~\bibnamefont {Vallone}}, \bibinfo {author} {\bibfnamefont {P.}~\bibnamefont {Villoresi}},\ and\ \bibinfo {author} {\bibfnamefont {P.}~\bibnamefont {Wallden}},\ }\bibfield  {title} {\bibinfo {title} {{Advances in quantum cryptography}},\ }\href {https://doi.org/10.1364/AOP.361502} {\bibfield  {journal} {\bibinfo  {journal} {Adv. Opt. Photonics}\ }\textbf {\bibinfo {volume} {12}},\ \bibinfo {pages} {1012} (\bibinfo {year} {2020})}\BibitemShut {NoStop}%
\bibitem [{\citenamefont {Agnesi}\ \emph {et~al.}(2020)\citenamefont {Agnesi}, \citenamefont {Avesani}, \citenamefont {Calderaro}, \citenamefont {Stanco}, \citenamefont {Foletto}, \citenamefont {Zahidy}, \citenamefont {Scriminich}, \citenamefont {Vedovato}, \citenamefont {Vallone},\ and\ \citenamefont {Villoresi}}]{Agnesi:20}%
  \BibitemOpen
  \bibfield  {author} {\bibinfo {author} {\bibfnamefont {C.}~\bibnamefont {Agnesi}}, \bibinfo {author} {\bibfnamefont {M.}~\bibnamefont {Avesani}}, \bibinfo {author} {\bibfnamefont {L.}~\bibnamefont {Calderaro}}, \bibinfo {author} {\bibfnamefont {A.}~\bibnamefont {Stanco}}, \bibinfo {author} {\bibfnamefont {G.}~\bibnamefont {Foletto}}, \bibinfo {author} {\bibfnamefont {M.}~\bibnamefont {Zahidy}}, \bibinfo {author} {\bibfnamefont {A.}~\bibnamefont {Scriminich}}, \bibinfo {author} {\bibfnamefont {F.}~\bibnamefont {Vedovato}}, \bibinfo {author} {\bibfnamefont {G.}~\bibnamefont {Vallone}},\ and\ \bibinfo {author} {\bibfnamefont {P.}~\bibnamefont {Villoresi}},\ }\bibfield  {title} {\bibinfo {title} {Simple quantum key distribution with qubit-based synchronization and a self-compensating polarization encoder},\ }\href {https://doi.org/10.1364/OPTICA.381013} {\bibfield  {journal} {\bibinfo  {journal} {Optica}\ }\textbf {\bibinfo {volume} {7}},\ \bibinfo {pages} {284} (\bibinfo {year} {2020})}\BibitemShut {NoStop}%
\bibitem [{\citenamefont {Bennett}(1992)}]{Bennett1992}%
  \BibitemOpen
  \bibfield  {author} {\bibinfo {author} {\bibfnamefont {C.~H.}\ \bibnamefont {Bennett}},\ }\bibfield  {title} {\bibinfo {title} {Quantum cryptography using any two nonorthogonal states},\ }\href {https://doi.org/10.1103/PhysRevLett.68.3121} {\bibfield  {journal} {\bibinfo  {journal} {Phys. Rev. Lett.}\ }\textbf {\bibinfo {volume} {68}},\ \bibinfo {pages} {3121} (\bibinfo {year} {1992})}\BibitemShut {NoStop}%
\bibitem [{\citenamefont {Ding}\ \emph {et~al.}(2017)\citenamefont {Ding}, \citenamefont {Chen}, \citenamefont {Wang}, \citenamefont {He}, \citenamefont {Yin}, \citenamefont {Chen}, \citenamefont {Zhou}, \citenamefont {Guo},\ and\ \citenamefont {Han}}]{Ding2017}%
  \BibitemOpen
  \bibfield  {author} {\bibinfo {author} {\bibfnamefont {Y.-Y.}\ \bibnamefont {Ding}}, \bibinfo {author} {\bibfnamefont {H.}~\bibnamefont {Chen}}, \bibinfo {author} {\bibfnamefont {S.}~\bibnamefont {Wang}}, \bibinfo {author} {\bibfnamefont {D.-Y.}\ \bibnamefont {He}}, \bibinfo {author} {\bibfnamefont {Z.-Q.}\ \bibnamefont {Yin}}, \bibinfo {author} {\bibfnamefont {W.}~\bibnamefont {Chen}}, \bibinfo {author} {\bibfnamefont {Z.}~\bibnamefont {Zhou}}, \bibinfo {author} {\bibfnamefont {G.-C.}\ \bibnamefont {Guo}},\ and\ \bibinfo {author} {\bibfnamefont {Z.-F.}\ \bibnamefont {Han}},\ }\bibfield  {title} {\bibinfo {title} {Polarization variations in installed fibers and their influence on quantum key distribution systems},\ }\href {https://doi.org/10.1364/OE.25.027923} {\bibfield  {journal} {\bibinfo  {journal} {Opt. Express}\ }\textbf {\bibinfo {volume} {25}},\ \bibinfo {pages} {27923} (\bibinfo {year} {2017})}\BibitemShut {NoStop}%
\bibitem [{\citenamefont {Li}\ \emph {et~al.}(2018)\citenamefont {Li}, \citenamefont {Gao}, \citenamefont {Li}, \citenamefont {Xue}, \citenamefont {Wang}, \citenamefont {Lu}, \citenamefont {Xiang}, \citenamefont {Zhao}, \citenamefont {Yan}, \citenamefont {Chen}, \citenamefont {Yu},\ and\ \citenamefont {Liu}}]{Li:18}%
  \BibitemOpen
  \bibfield  {author} {\bibinfo {author} {\bibfnamefont {D.-D.}\ \bibnamefont {Li}}, \bibinfo {author} {\bibfnamefont {S.}~\bibnamefont {Gao}}, \bibinfo {author} {\bibfnamefont {G.-C.}\ \bibnamefont {Li}}, \bibinfo {author} {\bibfnamefont {L.}~\bibnamefont {Xue}}, \bibinfo {author} {\bibfnamefont {L.-W.}\ \bibnamefont {Wang}}, \bibinfo {author} {\bibfnamefont {C.-B.}\ \bibnamefont {Lu}}, \bibinfo {author} {\bibfnamefont {Y.}~\bibnamefont {Xiang}}, \bibinfo {author} {\bibfnamefont {Z.-Y.}\ \bibnamefont {Zhao}}, \bibinfo {author} {\bibfnamefont {L.-C.}\ \bibnamefont {Yan}}, \bibinfo {author} {\bibfnamefont {Z.-Y.}\ \bibnamefont {Chen}}, \bibinfo {author} {\bibfnamefont {G.}~\bibnamefont {Yu}},\ and\ \bibinfo {author} {\bibfnamefont {J.-H.}\ \bibnamefont {Liu}},\ }\bibfield  {title} {\bibinfo {title} {Field implementation of long-distance quantum key distribution over aerial fiber with fast polarization feedback},\ }\href {https://doi.org/10.1364/OE.26.022793} {\bibfield  {journal} {\bibinfo  {journal} {Opt.
  Express}\ }\textbf {\bibinfo {volume} {26}},\ \bibinfo {pages} {22793} (\bibinfo {year} {2018})}\BibitemShut {NoStop}%
\bibitem [{\citenamefont {Avesani}\ \emph {et~al.}(2021)\citenamefont {Avesani}, \citenamefont {Calderaro}, \citenamefont {Foletto}, \citenamefont {Agnesi}, \citenamefont {Picciariello}, \citenamefont {Santagiustina}, \citenamefont {Scriminich}, \citenamefont {Stanco}, \citenamefont {Vedovato}, \citenamefont {Zahidy}, \citenamefont {Vallone},\ and\ \citenamefont {Villoresi}}]{Avesani:21}%
  \BibitemOpen
  \bibfield  {author} {\bibinfo {author} {\bibfnamefont {M.}~\bibnamefont {Avesani}}, \bibinfo {author} {\bibfnamefont {L.}~\bibnamefont {Calderaro}}, \bibinfo {author} {\bibfnamefont {G.}~\bibnamefont {Foletto}}, \bibinfo {author} {\bibfnamefont {C.}~\bibnamefont {Agnesi}}, \bibinfo {author} {\bibfnamefont {F.}~\bibnamefont {Picciariello}}, \bibinfo {author} {\bibfnamefont {F.~B.~L.}\ \bibnamefont {Santagiustina}}, \bibinfo {author} {\bibfnamefont {A.}~\bibnamefont {Scriminich}}, \bibinfo {author} {\bibfnamefont {A.}~\bibnamefont {Stanco}}, \bibinfo {author} {\bibfnamefont {F.}~\bibnamefont {Vedovato}}, \bibinfo {author} {\bibfnamefont {M.}~\bibnamefont {Zahidy}}, \bibinfo {author} {\bibfnamefont {G.}~\bibnamefont {Vallone}},\ and\ \bibinfo {author} {\bibfnamefont {P.}~\bibnamefont {Villoresi}},\ }\bibfield  {title} {\bibinfo {title} {Resource-effective quantum key distribution: a field trial in {Padua} city center},\ }\href {https://doi.org/10.1364/OL.422890} {\bibfield  {journal} {\bibinfo  {journal} {Opt.
  Lett.}\ }\textbf {\bibinfo {volume} {46}},\ \bibinfo {pages} {2848} (\bibinfo {year} {2021})}\BibitemShut {NoStop}%
\bibitem [{\citenamefont {Agnesi}\ \emph {et~al.}(2024)\citenamefont {Agnesi}, \citenamefont {Giacomin}, \citenamefont {Sartorato}, \citenamefont {Artuso}, \citenamefont {Vallone},\ and\ \citenamefont {Villoresi}}]{Telebit}%
  \BibitemOpen
  \bibfield  {author} {\bibinfo {author} {\bibfnamefont {C.}~\bibnamefont {Agnesi}}, \bibinfo {author} {\bibfnamefont {M.}~\bibnamefont {Giacomin}}, \bibinfo {author} {\bibfnamefont {D.}~\bibnamefont {Sartorato}}, \bibinfo {author} {\bibfnamefont {S.}~\bibnamefont {Artuso}}, \bibinfo {author} {\bibfnamefont {G.}~\bibnamefont {Vallone}},\ and\ \bibinfo {author} {\bibfnamefont {P.}~\bibnamefont {Villoresi}},\ }\bibfield  {title} {\bibinfo {title} {In-field comparison between {G.652} and {G.655} optical fibres for polarisation-based quantum key distribution},\ }\href {https://doi.org/https://doi.org/10.1049/qtc2.12095} {\bibfield  {journal} {\bibinfo  {journal} {IET Quantum Comm.}\ ,\ \bibinfo {pages} {1}} (\bibinfo {year} {2024})}\BibitemShut {NoStop}%
\bibitem [{\citenamefont {Makarov}\ \emph {et~al.}(2004)\citenamefont {Makarov}, \citenamefont {Brylevski},\ and\ \citenamefont {Hjelme}}]{Makarov:04}%
  \BibitemOpen
  \bibfield  {author} {\bibinfo {author} {\bibfnamefont {V.}~\bibnamefont {Makarov}}, \bibinfo {author} {\bibfnamefont {A.}~\bibnamefont {Brylevski}},\ and\ \bibinfo {author} {\bibfnamefont {D.~R.}\ \bibnamefont {Hjelme}},\ }\bibfield  {title} {\bibinfo {title} {Real-time phase tracking in single-photon interferometers},\ }\href {https://doi.org/10.1364/AO.43.004385} {\bibfield  {journal} {\bibinfo  {journal} {Appl. Opt.}\ }\textbf {\bibinfo {volume} {43}},\ \bibinfo {pages} {4385} (\bibinfo {year} {2004})}\BibitemShut {NoStop}%
\bibitem [{\citenamefont {\v{S}varc}\ \emph {et~al.}(2023)\citenamefont {\v{S}varc}, \citenamefont {Nov\'{a}kov\'{a}}, \citenamefont {Dudka},\ and\ \citenamefont {Je\v{z}ek}}]{Svarc:23}%
  \BibitemOpen
  \bibfield  {author} {\bibinfo {author} {\bibfnamefont {V.}~\bibnamefont {\v{S}varc}}, \bibinfo {author} {\bibfnamefont {M.}~\bibnamefont {Nov\'{a}kov\'{a}}}, \bibinfo {author} {\bibfnamefont {M.}~\bibnamefont {Dudka}},\ and\ \bibinfo {author} {\bibfnamefont {M.}~\bibnamefont {Je\v{z}ek}},\ }\bibfield  {title} {\bibinfo {title} {Sub-0.1 degree phase locking of a single-photon interferometer},\ }\href {https://doi.org/10.1364/OE.480569} {\bibfield  {journal} {\bibinfo  {journal} {Opt. Express}\ }\textbf {\bibinfo {volume} {31}},\ \bibinfo {pages} {12562} (\bibinfo {year} {2023})}\BibitemShut {NoStop}%
\bibitem [{\citenamefont {Hacker}\ \emph {et~al.}(2023)\citenamefont {Hacker}, \citenamefont {Günthner}, \citenamefont {Rößler},\ and\ \citenamefont {Marquardt}}]{Hacker_2023}%
  \BibitemOpen
  \bibfield  {author} {\bibinfo {author} {\bibfnamefont {B.}~\bibnamefont {Hacker}}, \bibinfo {author} {\bibfnamefont {K.}~\bibnamefont {Günthner}}, \bibinfo {author} {\bibfnamefont {C.}~\bibnamefont {Rößler}},\ and\ \bibinfo {author} {\bibfnamefont {C.}~\bibnamefont {Marquardt}},\ }\bibfield  {title} {\bibinfo {title} {Phase-locking an interferometer with single-photon detections},\ }\href {https://doi.org/10.1088/1367-2630/ad0752} {\bibfield  {journal} {\bibinfo  {journal} {New J. Phys.}\ }\textbf {\bibinfo {volume} {25}},\ \bibinfo {pages} {113007} (\bibinfo {year} {2023})}\BibitemShut {NoStop}%
\bibitem [{\citenamefont {Laing}\ \emph {et~al.}(2010)\citenamefont {Laing}, \citenamefont {Scarani}, \citenamefont {Rarity},\ and\ \citenamefont {O'Brien}}]{originalRFI2010}%
  \BibitemOpen
  \bibfield  {author} {\bibinfo {author} {\bibfnamefont {A.}~\bibnamefont {Laing}}, \bibinfo {author} {\bibfnamefont {V.}~\bibnamefont {Scarani}}, \bibinfo {author} {\bibfnamefont {J.~G.}\ \bibnamefont {Rarity}},\ and\ \bibinfo {author} {\bibfnamefont {J.~L.}\ \bibnamefont {O'Brien}},\ }\bibfield  {title} {\bibinfo {title} {Reference-frame-independent quantum key distribution},\ }\href {https://doi.org/10.1103/PhysRevA.82.012304} {\bibfield  {journal} {\bibinfo  {journal} {Phys. Rev. A}\ }\textbf {\bibinfo {volume} {82}},\ \bibinfo {pages} {012304} (\bibinfo {year} {2010})}\BibitemShut {NoStop}%
\bibitem [{\citenamefont {Wang}\ \emph {et~al.}(2015)\citenamefont {Wang}, \citenamefont {Sun}, \citenamefont {Ma}, \citenamefont {Tang},\ and\ \citenamefont {Liang}}]{previous_rfi_2}%
  \BibitemOpen
  \bibfield  {author} {\bibinfo {author} {\bibfnamefont {C.}~\bibnamefont {Wang}}, \bibinfo {author} {\bibfnamefont {S.-H.}\ \bibnamefont {Sun}}, \bibinfo {author} {\bibfnamefont {X.-C.}\ \bibnamefont {Ma}}, \bibinfo {author} {\bibfnamefont {G.-Z.}\ \bibnamefont {Tang}},\ and\ \bibinfo {author} {\bibfnamefont {L.-M.}\ \bibnamefont {Liang}},\ }\bibfield  {title} {\bibinfo {title} {Reference-frame-independent quantum key distribution with source flaws},\ }\href {https://doi.org/10.1103/PhysRevA.92.042319} {\bibfield  {journal} {\bibinfo  {journal} {Phys. Rev. A}\ }\textbf {\bibinfo {volume} {92}},\ \bibinfo {pages} {042319} (\bibinfo {year} {2015})}\BibitemShut {NoStop}%
\bibitem [{\citenamefont {Liu}\ \emph {et~al.}(2019)\citenamefont {Liu}, \citenamefont {Wang}, \citenamefont {Ma},\ and\ \citenamefont {Sun}}]{RFI_QKD_china}%
  \BibitemOpen
  \bibfield  {author} {\bibinfo {author} {\bibfnamefont {H.}~\bibnamefont {Liu}}, \bibinfo {author} {\bibfnamefont {J.}~\bibnamefont {Wang}}, \bibinfo {author} {\bibfnamefont {H.}~\bibnamefont {Ma}},\ and\ \bibinfo {author} {\bibfnamefont {S.}~\bibnamefont {Sun}},\ }\bibfield  {title} {\bibinfo {title} {Reference-frame-independent quantum key distribution using fewer states},\ }\href {https://doi.org/10.1103/PhysRevApplied.12.034039} {\bibfield  {journal} {\bibinfo  {journal} {Phys. Rev. Applied}\ }\textbf {\bibinfo {volume} {12}},\ \bibinfo {pages} {034039} (\bibinfo {year} {2019})}\BibitemShut {NoStop}%
\bibitem [{\citenamefont {Wang}\ \emph {et~al.}(2019)\citenamefont {Wang}, \citenamefont {Liu}, \citenamefont {Ma},\ and\ \citenamefont {Sun}}]{Wang2019}%
  \BibitemOpen
  \bibfield  {author} {\bibinfo {author} {\bibfnamefont {J.}~\bibnamefont {Wang}}, \bibinfo {author} {\bibfnamefont {H.}~\bibnamefont {Liu}}, \bibinfo {author} {\bibfnamefont {H.}~\bibnamefont {Ma}},\ and\ \bibinfo {author} {\bibfnamefont {S.}~\bibnamefont {Sun}},\ }\bibfield  {title} {\bibinfo {title} {Experimental study of four-state reference-frame-independent quantum key distribution with source flaws},\ }\href {https://doi.org/10.1103/PhysRevA.99.032309} {\bibfield  {journal} {\bibinfo  {journal} {Phys. Rev. A}\ }\textbf {\bibinfo {volume} {99}},\ \bibinfo {pages} {032309} (\bibinfo {year} {2019})}\BibitemShut {NoStop}%
\bibitem [{\citenamefont {Chen}\ \emph {et~al.}(2020)\citenamefont {Chen}, \citenamefont {Wang}, \citenamefont {Tang}, \citenamefont {Li}, \citenamefont {Liu},\ and\ \citenamefont {Sun}}]{Chen2020}%
  \BibitemOpen
  \bibfield  {author} {\bibinfo {author} {\bibfnamefont {H.}~\bibnamefont {Chen}}, \bibinfo {author} {\bibfnamefont {J.}~\bibnamefont {Wang}}, \bibinfo {author} {\bibfnamefont {B.}~\bibnamefont {Tang}}, \bibinfo {author} {\bibfnamefont {Z.}~\bibnamefont {Li}}, \bibinfo {author} {\bibfnamefont {B.}~\bibnamefont {Liu}},\ and\ \bibinfo {author} {\bibfnamefont {S.}~\bibnamefont {Sun}},\ }\bibfield  {title} {\bibinfo {title} {Field demonstration of time-bin reference-frame-independent quantum key distribution via an intracity free-space link},\ }\href {https://doi.org/10.1364/OL.392742} {\bibfield  {journal} {\bibinfo  {journal} {Opt. Lett.}\ }\textbf {\bibinfo {volume} {45}},\ \bibinfo {pages} {3022} (\bibinfo {year} {2020})}\BibitemShut {NoStop}%
\bibitem [{\citenamefont {Tang}\ \emph {et~al.}(2022)\citenamefont {Tang}, \citenamefont {Chen}, \citenamefont {Wang}, \citenamefont {Yu}, \citenamefont {Shi}, \citenamefont {Sun}, \citenamefont {Peng}, \citenamefont {Liu},\ and\ \citenamefont {Yu}}]{Tang2022}%
  \BibitemOpen
  \bibfield  {author} {\bibinfo {author} {\bibfnamefont {B.-Y.}\ \bibnamefont {Tang}}, \bibinfo {author} {\bibfnamefont {H.}~\bibnamefont {Chen}}, \bibinfo {author} {\bibfnamefont {J.-P.}\ \bibnamefont {Wang}}, \bibinfo {author} {\bibfnamefont {H.-C.}\ \bibnamefont {Yu}}, \bibinfo {author} {\bibfnamefont {L.}~\bibnamefont {Shi}}, \bibinfo {author} {\bibfnamefont {S.-H.}\ \bibnamefont {Sun}}, \bibinfo {author} {\bibfnamefont {W.}~\bibnamefont {Peng}}, \bibinfo {author} {\bibfnamefont {B.}~\bibnamefont {Liu}},\ and\ \bibinfo {author} {\bibfnamefont {W.-R.}\ \bibnamefont {Yu}},\ }\bibfield  {title} {\bibinfo {title} {Free-running long-distance reference-frame-independent quantum key distribution},\ }\href {https://doi.org/10.1038/s41534-022-00630-3} {\bibfield  {journal} {\bibinfo  {journal} {npj Quantum Inf.}\ }\textbf {\bibinfo {volume} {8}},\ \bibinfo {pages} {117} (\bibinfo {year} {2022})}\BibitemShut {NoStop}%
\bibitem [{\citenamefont {Tannous}\ \emph {et~al.}(2023)\citenamefont {Tannous}, \citenamefont {Wu}, \citenamefont {Vinet}, \citenamefont {Perumangatt}, \citenamefont {Sinar}, \citenamefont {Ling},\ and\ \citenamefont {Jennewein}}]{RFI_waterloo}%
  \BibitemOpen
  \bibfield  {author} {\bibinfo {author} {\bibfnamefont {R.}~\bibnamefont {Tannous}}, \bibinfo {author} {\bibfnamefont {W.}~\bibnamefont {Wu}}, \bibinfo {author} {\bibfnamefont {S.}~\bibnamefont {Vinet}}, \bibinfo {author} {\bibfnamefont {C.}~\bibnamefont {Perumangatt}}, \bibinfo {author} {\bibfnamefont {D.}~\bibnamefont {Sinar}}, \bibinfo {author} {\bibfnamefont {A.}~\bibnamefont {Ling}},\ and\ \bibinfo {author} {\bibfnamefont {T.}~\bibnamefont {Jennewein}},\ }\bibfield  {title} {\bibinfo {title} {Towards fully passive time-bin quantum key distribution over multi-mode channels}\ }\href {https://doi.org/arXiv:2302.05038} {arXiv:2302.05038} (\bibinfo {year} {2023})\BibitemShut {NoStop}%
\bibitem [{\citenamefont {Jennewein}\ \emph {et~al.}(2000)\citenamefont {Jennewein}, \citenamefont {Achleitner}, \citenamefont {Weihs}, \citenamefont {Weinfurter},\ and\ \citenamefont {Zeilinger}}]{Jennewein2000}%
  \BibitemOpen
  \bibfield  {author} {\bibinfo {author} {\bibfnamefont {T.}~\bibnamefont {Jennewein}}, \bibinfo {author} {\bibfnamefont {U.}~\bibnamefont {Achleitner}}, \bibinfo {author} {\bibfnamefont {G.}~\bibnamefont {Weihs}}, \bibinfo {author} {\bibfnamefont {H.}~\bibnamefont {Weinfurter}},\ and\ \bibinfo {author} {\bibfnamefont {A.}~\bibnamefont {Zeilinger}},\ }\bibfield  {title} {\bibinfo {title} {{A fast and compact quantum random number generator}},\ }\href {https://doi.org/10.1063/1.1150518} {\bibfield  {journal} {\bibinfo  {journal} {Rev. Sci. Instrum.}\ }\textbf {\bibinfo {volume} {71}},\ \bibinfo {pages} {1675} (\bibinfo {year} {2000})}\BibitemShut {NoStop}%
\bibitem [{\citenamefont {Stipčević}\ and\ \citenamefont {Rogina}(2007)}]{Stipcevic2007}%
  \BibitemOpen
  \bibfield  {author} {\bibinfo {author} {\bibfnamefont {M.}~\bibnamefont {Stipčević}}\ and\ \bibinfo {author} {\bibfnamefont {B.~M.}\ \bibnamefont {Rogina}},\ }\bibfield  {title} {\bibinfo {title} {{Quantum random number generator based on photonic emission in semiconductors}},\ }\href {https://doi.org/10.1063/1.2720728} {\bibfield  {journal} {\bibinfo  {journal} {Rev. Sci. Instrum.}\ }\textbf {\bibinfo {volume} {78}},\ \bibinfo {pages} {045104} (\bibinfo {year} {2007})}\BibitemShut {NoStop}%
\bibitem [{\citenamefont {Stanco}\ \emph {et~al.}(2020)\citenamefont {Stanco}, \citenamefont {Marangon}, \citenamefont {Vallone}, \citenamefont {Burri}, \citenamefont {Charbon},\ and\ \citenamefont {Villoresi}}]{QRNG2020}%
  \BibitemOpen
  \bibfield  {author} {\bibinfo {author} {\bibfnamefont {A.}~\bibnamefont {Stanco}}, \bibinfo {author} {\bibfnamefont {D.~G.}\ \bibnamefont {Marangon}}, \bibinfo {author} {\bibfnamefont {G.}~\bibnamefont {Vallone}}, \bibinfo {author} {\bibfnamefont {S.}~\bibnamefont {Burri}}, \bibinfo {author} {\bibfnamefont {E.}~\bibnamefont {Charbon}},\ and\ \bibinfo {author} {\bibfnamefont {P.}~\bibnamefont {Villoresi}},\ }\bibfield  {title} {\bibinfo {title} {Efficient random number generation techniques for {CMOS} single-photon avalanche diode array exploiting fast time tagging units},\ }\href {https://doi.org/10.1103/PhysRevResearch.2.023287} {\bibfield  {journal} {\bibinfo  {journal} {Phys. Rev. Res.}\ }\textbf {\bibinfo {volume} {2}},\ \bibinfo {pages} {023287} (\bibinfo {year} {2020})}\BibitemShut {NoStop}%
\bibitem [{\citenamefont {Stanco}\ \emph {et~al.}(2022)\citenamefont {Stanco}, \citenamefont {Santagiustina}, \citenamefont {Calderaro}, \citenamefont {Avesani}, \citenamefont {Bertapelle}, \citenamefont {Dequal}, \citenamefont {Vallone},\ and\ \citenamefont {Villoresi}}]{Stanco2022}%
  \BibitemOpen
  \bibfield  {author} {\bibinfo {author} {\bibfnamefont {A.}~\bibnamefont {Stanco}}, \bibinfo {author} {\bibfnamefont {F.~B.~L.}\ \bibnamefont {Santagiustina}}, \bibinfo {author} {\bibfnamefont {L.}~\bibnamefont {Calderaro}}, \bibinfo {author} {\bibfnamefont {M.}~\bibnamefont {Avesani}}, \bibinfo {author} {\bibfnamefont {T.}~\bibnamefont {Bertapelle}}, \bibinfo {author} {\bibfnamefont {D.}~\bibnamefont {Dequal}}, \bibinfo {author} {\bibfnamefont {G.}~\bibnamefont {Vallone}},\ and\ \bibinfo {author} {\bibfnamefont {P.}~\bibnamefont {Villoresi}},\ }\bibfield  {title} {\bibinfo {title} {Versatile and concurrent fpga-based architecture for practical quantum communication systems},\ }\href {https://doi.org/10.1109/TQE.2022.3143997} {\bibfield  {journal} {\bibinfo  {journal} {IEEE Trans. Quantum Eng.}\ }\textbf {\bibinfo {volume} {3}},\ \bibinfo {pages} {6000108} (\bibinfo {year} {2022})}\BibitemShut {NoStop}%
\bibitem [{\citenamefont {Scalcon}\ \emph {et~al.}(2022)\citenamefont {Scalcon}, \citenamefont {Agnesi}, \citenamefont {Avesani}, \citenamefont {Calderaro}, \citenamefont {Foletto}, \citenamefont {Stanco}, \citenamefont {Vallone},\ and\ \citenamefont {Villoresi}}]{cross-encoding}%
  \BibitemOpen
  \bibfield  {author} {\bibinfo {author} {\bibfnamefont {D.}~\bibnamefont {Scalcon}}, \bibinfo {author} {\bibfnamefont {C.}~\bibnamefont {Agnesi}}, \bibinfo {author} {\bibfnamefont {M.}~\bibnamefont {Avesani}}, \bibinfo {author} {\bibfnamefont {L.}~\bibnamefont {Calderaro}}, \bibinfo {author} {\bibfnamefont {G.}~\bibnamefont {Foletto}}, \bibinfo {author} {\bibfnamefont {A.}~\bibnamefont {Stanco}}, \bibinfo {author} {\bibfnamefont {G.}~\bibnamefont {Vallone}},\ and\ \bibinfo {author} {\bibfnamefont {P.}~\bibnamefont {Villoresi}},\ }\bibfield  {title} {\bibinfo {title} {Cross-encoded quantum key distribution exploiting time-bin and polarization states with qubit-based synchronization},\ }\href {https://doi.org/https://doi.org/10.1002/qute.202200051} {\bibfield  {journal} {\bibinfo  {journal} {Adv. Quantum Technol.}\ }\textbf {\bibinfo {volume} {5}},\ \bibinfo {pages} {2200051} (\bibinfo {year} {2022})}\BibitemShut {NoStop}%
\bibitem [{\citenamefont {Zhang}\ \emph {et~al.}(2020)\citenamefont {Zhang}, \citenamefont {Xie}, \citenamefont {Xu}, \citenamefont {Zheng}, \citenamefont {Zhang}, \citenamefont {Poon}, \citenamefont {Vedral},\ and\ \citenamefont {Zhang}}]{QDSC}%
  \BibitemOpen
  \bibfield  {author} {\bibinfo {author} {\bibfnamefont {A.}~\bibnamefont {Zhang}}, \bibinfo {author} {\bibfnamefont {J.}~\bibnamefont {Xie}}, \bibinfo {author} {\bibfnamefont {H.}~\bibnamefont {Xu}}, \bibinfo {author} {\bibfnamefont {K.}~\bibnamefont {Zheng}}, \bibinfo {author} {\bibfnamefont {H.}~\bibnamefont {Zhang}}, \bibinfo {author} {\bibfnamefont {Y.-T.}\ \bibnamefont {Poon}}, \bibinfo {author} {\bibfnamefont {V.}~\bibnamefont {Vedral}},\ and\ \bibinfo {author} {\bibfnamefont {L.}~\bibnamefont {Zhang}},\ }\bibfield  {title} {\bibinfo {title} {Experimental self-characterization of quantum measurements},\ }\href {https://doi.org/10.1103/PhysRevLett.124.040402} {\bibfield  {journal} {\bibinfo  {journal} {Phys. Rev. Lett.}\ }\textbf {\bibinfo {volume} {124}},\ \bibinfo {pages} {040402} (\bibinfo {year} {2020})}\BibitemShut {NoStop}%
\bibitem [{\citenamefont {Liao}\ \emph {et~al.}(2017)\citenamefont {Liao}, \citenamefont {Cai}, \citenamefont {Liu}, \citenamefont {Zhang}, \citenamefont {Li}, \citenamefont {Ren}, \citenamefont {Yin}, \citenamefont {Shen}, \citenamefont {Cao}, \citenamefont {Li}, \citenamefont {Li}, \citenamefont {Chen}, \citenamefont {Sun}, \citenamefont {Jia}, \citenamefont {Wu}, \citenamefont {Jiang}, \citenamefont {Wang}, \citenamefont {Huang}, \citenamefont {Wang}, \citenamefont {Zhou}, \citenamefont {Deng}, \citenamefont {Xi}, \citenamefont {Ma}, \citenamefont {Hu}, \citenamefont {Zhang}, \citenamefont {Chen}, \citenamefont {Liu}, \citenamefont {Wang}, \citenamefont {Zhu}, \citenamefont {Lu}, \citenamefont {Shu}, \citenamefont {Peng}, \citenamefont {Wang},\ and\ \citenamefont {Pan}}]{Liao2017_Sat}%
  \BibitemOpen
  \bibfield  {author} {\bibinfo {author} {\bibfnamefont {S.-K.}\ \bibnamefont {Liao}}, \bibinfo {author} {\bibfnamefont {W.-Q.}\ \bibnamefont {Cai}}, \bibinfo {author} {\bibfnamefont {W.-Y.}\ \bibnamefont {Liu}}, \bibinfo {author} {\bibfnamefont {L.}~\bibnamefont {Zhang}}, \bibinfo {author} {\bibfnamefont {Y.}~\bibnamefont {Li}}, \bibinfo {author} {\bibfnamefont {J.-G.}\ \bibnamefont {Ren}}, \bibinfo {author} {\bibfnamefont {J.}~\bibnamefont {Yin}}, \bibinfo {author} {\bibfnamefont {Q.}~\bibnamefont {Shen}}, \bibinfo {author} {\bibfnamefont {Y.}~\bibnamefont {Cao}}, \bibinfo {author} {\bibfnamefont {Z.-P.}\ \bibnamefont {Li}}, \bibinfo {author} {\bibfnamefont {F.-Z.}\ \bibnamefont {Li}}, \bibinfo {author} {\bibfnamefont {X.-W.}\ \bibnamefont {Chen}}, \bibinfo {author} {\bibfnamefont {L.-H.}\ \bibnamefont {Sun}}, \bibinfo {author} {\bibfnamefont {J.-J.}\ \bibnamefont {Jia}}, \bibinfo {author} {\bibfnamefont {J.-C.}\ \bibnamefont {Wu}}, \bibinfo {author} {\bibfnamefont {X.-J.}\ \bibnamefont {Jiang}}, \bibinfo
  {author} {\bibfnamefont {J.-F.}\ \bibnamefont {Wang}}, \bibinfo {author} {\bibfnamefont {Y.-M.}\ \bibnamefont {Huang}}, \bibinfo {author} {\bibfnamefont {Q.}~\bibnamefont {Wang}}, \bibinfo {author} {\bibfnamefont {Y.-L.}\ \bibnamefont {Zhou}}, \bibinfo {author} {\bibfnamefont {L.}~\bibnamefont {Deng}}, \bibinfo {author} {\bibfnamefont {T.}~\bibnamefont {Xi}}, \bibinfo {author} {\bibfnamefont {L.}~\bibnamefont {Ma}}, \bibinfo {author} {\bibfnamefont {T.}~\bibnamefont {Hu}}, \bibinfo {author} {\bibfnamefont {Q.}~\bibnamefont {Zhang}}, \bibinfo {author} {\bibfnamefont {Y.-A.}\ \bibnamefont {Chen}}, \bibinfo {author} {\bibfnamefont {N.-L.}\ \bibnamefont {Liu}}, \bibinfo {author} {\bibfnamefont {X.-B.}\ \bibnamefont {Wang}}, \bibinfo {author} {\bibfnamefont {Z.-C.}\ \bibnamefont {Zhu}}, \bibinfo {author} {\bibfnamefont {C.-Y.}\ \bibnamefont {Lu}}, \bibinfo {author} {\bibfnamefont {R.}~\bibnamefont {Shu}}, \bibinfo {author} {\bibfnamefont {C.-Z.}\ \bibnamefont {Peng}}, \bibinfo {author} {\bibfnamefont {J.-Y.}\
  \bibnamefont {Wang}},\ and\ \bibinfo {author} {\bibfnamefont {J.-W.}\ \bibnamefont {Pan}},\ }\bibfield  {title} {\bibinfo {title} {{Satellite-to-ground quantum key distribution}},\ }\href {https://doi.org/10.1038/nature23655} {\bibfield  {journal} {\bibinfo  {journal} {Nature}\ }\textbf {\bibinfo {volume} {549}},\ \bibinfo {pages} {43} (\bibinfo {year} {2017})}\BibitemShut {NoStop}%
\bibitem [{\citenamefont {Berra}\ \emph {et~al.}(2023)\citenamefont {Berra}, \citenamefont {Agnesi}, \citenamefont {Stanco}, \citenamefont {Avesani}, \citenamefont {Cocchi}, \citenamefont {Villoresi},\ and\ \citenamefont {Vallone}}]{Berra2023}%
  \BibitemOpen
  \bibfield  {author} {\bibinfo {author} {\bibfnamefont {F.}~\bibnamefont {Berra}}, \bibinfo {author} {\bibfnamefont {C.}~\bibnamefont {Agnesi}}, \bibinfo {author} {\bibfnamefont {A.}~\bibnamefont {Stanco}}, \bibinfo {author} {\bibfnamefont {M.}~\bibnamefont {Avesani}}, \bibinfo {author} {\bibfnamefont {S.}~\bibnamefont {Cocchi}}, \bibinfo {author} {\bibfnamefont {P.}~\bibnamefont {Villoresi}},\ and\ \bibinfo {author} {\bibfnamefont {G.}~\bibnamefont {Vallone}},\ }\bibfield  {title} {\bibinfo {title} {Modular source for near-infrared quantum communication},\ }\href {https://doi.org/10.1140/epjqt/s40507-023-00185-y} {\bibfield  {journal} {\bibinfo  {journal} {EPJ Quantum Technol.}\ }\textbf {\bibinfo {volume} {10}},\ \bibinfo {pages} {27} (\bibinfo {year} {2023})}\BibitemShut {NoStop}%
\bibitem [{\citenamefont {Dall'Arno}\ \emph {et~al.}(2017)\citenamefont {Dall'Arno}, \citenamefont {Brandsen}, \citenamefont {Buscemi},\ and\ \citenamefont {Vedral}}]{DallArno2017}%
  \BibitemOpen
  \bibfield  {author} {\bibinfo {author} {\bibfnamefont {M.}~\bibnamefont {Dall'Arno}}, \bibinfo {author} {\bibfnamefont {S.}~\bibnamefont {Brandsen}}, \bibinfo {author} {\bibfnamefont {F.}~\bibnamefont {Buscemi}},\ and\ \bibinfo {author} {\bibfnamefont {V.}~\bibnamefont {Vedral}},\ }\bibfield  {title} {\bibinfo {title} {Device-independent tests of quantum measurements},\ }\href {https://link.aps.org/doi/10.1103/PhysRevLett.118.250501} {\bibfield  {journal} {\bibinfo  {journal} {Phys. Rev. Lett.}\ }\textbf {\bibinfo {volume} {118}},\ \bibinfo {pages} {250501} (\bibinfo {year} {2017})}\BibitemShut {NoStop}%
\bibitem [{\citenamefont {Jolliffe}\ and\ \citenamefont {Cadima}(2016)}]{doi:10.1098/rsta.2015.0202}%
  \BibitemOpen
  \bibfield  {author} {\bibinfo {author} {\bibfnamefont {I.~T.}\ \bibnamefont {Jolliffe}}\ and\ \bibinfo {author} {\bibfnamefont {J.}~\bibnamefont {Cadima}},\ }\bibfield  {title} {\bibinfo {title} {Principal component analysis: a review and recent developments},\ }\href {https://doi.org/10.1098/rsta.2015.0202} {\bibfield  {journal} {\bibinfo  {journal} {Philos. Trans. R. Soc. A}\ }\textbf {\bibinfo {volume} {374}},\ \bibinfo {pages} {20150202} (\bibinfo {year} {2016})}\BibitemShut {NoStop}%
\bibitem [{\citenamefont {Tamaki}\ \emph {et~al.}(2014)\citenamefont {Tamaki}, \citenamefont {Curty}, \citenamefont {Kato}, \citenamefont {Lo},\ and\ \citenamefont {Azuma}}]{previous_rfi_1}%
  \BibitemOpen
  \bibfield  {author} {\bibinfo {author} {\bibfnamefont {K.}~\bibnamefont {Tamaki}}, \bibinfo {author} {\bibfnamefont {M.}~\bibnamefont {Curty}}, \bibinfo {author} {\bibfnamefont {G.}~\bibnamefont {Kato}}, \bibinfo {author} {\bibfnamefont {H.-K.}\ \bibnamefont {Lo}},\ and\ \bibinfo {author} {\bibfnamefont {K.}~\bibnamefont {Azuma}},\ }\bibfield  {title} {\bibinfo {title} {Loss-tolerant quantum cryptography with imperfect sources},\ }\href {https://doi.org/10.1103/PhysRevA.90.052314} {\bibfield  {journal} {\bibinfo  {journal} {Phys. Rev. A}\ }\textbf {\bibinfo {volume} {90}},\ \bibinfo {pages} {052314} (\bibinfo {year} {2014})}\BibitemShut {NoStop}%
\bibitem [{\citenamefont {Sheridan}\ \emph {et~al.}(2010)\citenamefont {Sheridan}, \citenamefont {Le},\ and\ \citenamefont {Scarani}}]{Sheridan2010}%
  \BibitemOpen
  \bibfield  {author} {\bibinfo {author} {\bibfnamefont {L.}~\bibnamefont {Sheridan}}, \bibinfo {author} {\bibfnamefont {T.~P.}\ \bibnamefont {Le}},\ and\ \bibinfo {author} {\bibfnamefont {V.}~\bibnamefont {Scarani}},\ }\bibfield  {title} {\bibinfo {title} {Finite-key security against coherent attacks in quantum key distribution},\ }\href {https://doi.org/10.1088/1367-2630/12/12/123019} {\bibfield  {journal} {\bibinfo  {journal} {New J. Phys.}\ }\textbf {\bibinfo {volume} {12}},\ \bibinfo {pages} {123019} (\bibinfo {year} {2010})}\BibitemShut {NoStop}%
\bibitem [{\citenamefont {Grant}\ and\ \citenamefont {Boyd}(2014)}]{cvx_1}%
  \BibitemOpen
  \bibfield  {author} {\bibinfo {author} {\bibfnamefont {M.}~\bibnamefont {Grant}}\ and\ \bibinfo {author} {\bibfnamefont {S.}~\bibnamefont {Boyd}},\ }\href@noop {} {\bibinfo {title} {{CVX}: Matlab software for disciplined convex programming, version 2.1}},\ \bibinfo {howpublished} {\url{http://cvxr.com/cvx}} (\bibinfo {year} {2014})\BibitemShut {NoStop}%
\bibitem [{\citenamefont {Grant}\ and\ \citenamefont {Boyd}(2008)}]{cvx_2}%
  \BibitemOpen
  \bibfield  {author} {\bibinfo {author} {\bibfnamefont {M.}~\bibnamefont {Grant}}\ and\ \bibinfo {author} {\bibfnamefont {S.}~\bibnamefont {Boyd}},\ }\bibfield  {title} {\bibinfo {title} {Graph implementations for nonsmooth convex programs},\ }in\ \href@noop {} {\emph {\bibinfo {booktitle} {Recent Advances in Learning and Control}}},\ \bibinfo {series and number} {Lecture Notes in Control and Information Sciences},\ \bibinfo {editor} {edited by\ \bibinfo {editor} {\bibfnamefont {V.}~\bibnamefont {Blondel}}, \bibinfo {editor} {\bibfnamefont {S.}~\bibnamefont {Boyd}},\ and\ \bibinfo {editor} {\bibfnamefont {H.}~\bibnamefont {Kimura}}}\ (\bibinfo  {publisher} {Springer-Verlag Limited},\ \bibinfo {year} {2008})\ pp.\ \bibinfo {pages} {95--110},\ \bibinfo {note} {\url{http://stanford.edu/~boyd/graph_dcp.html}}\BibitemShut {NoStop}%
\bibitem [{\citenamefont {Sasaki}\ \emph {et~al.}(2011)\citenamefont {Sasaki}, \citenamefont {Fujiwara}, \citenamefont {Ishizuka}, \citenamefont {Klaus}, \citenamefont {Wakui}, \citenamefont {Takeoka}, \citenamefont {Miki}, \citenamefont {Yamashita}, \citenamefont {Wang}, \citenamefont {Tanaka}, \citenamefont {Yoshino}, \citenamefont {Nambu}, \citenamefont {Takahashi}, \citenamefont {Tajima}, \citenamefont {Tomita}, \citenamefont {Domeki}, \citenamefont {Hasegawa}, \citenamefont {Sakai}, \citenamefont {Kobayashi}, \citenamefont {Asai}, \citenamefont {Shimizu}, \citenamefont {Tokura}, \citenamefont {Tsurumaru}, \citenamefont {Matsui}, \citenamefont {Honjo}, \citenamefont {Tamaki}, \citenamefont {Takesue}, \citenamefont {Tokura}, \citenamefont {Dynes}, \citenamefont {Dixon}, \citenamefont {Sharpe}, \citenamefont {Yuan}, \citenamefont {Shields}, \citenamefont {Uchikoga}, \citenamefont {Legr\'{e}}, \citenamefont {Robyr}, \citenamefont {Trinkler}, \citenamefont {Monat}, \citenamefont {Page}, \citenamefont
  {Ribordy}, \citenamefont {Poppe}, \citenamefont {Allacher}, \citenamefont {Maurhart}, \citenamefont {L\"{a}nger}, \citenamefont {Peev},\ and\ \citenamefont {Zeilinger}}]{Sasaki:11}%
  \BibitemOpen
  \bibfield  {author} {\bibinfo {author} {\bibfnamefont {M.}~\bibnamefont {Sasaki}}, \bibinfo {author} {\bibfnamefont {M.}~\bibnamefont {Fujiwara}}, \bibinfo {author} {\bibfnamefont {H.}~\bibnamefont {Ishizuka}}, \bibinfo {author} {\bibfnamefont {W.}~\bibnamefont {Klaus}}, \bibinfo {author} {\bibfnamefont {K.}~\bibnamefont {Wakui}}, \bibinfo {author} {\bibfnamefont {M.}~\bibnamefont {Takeoka}}, \bibinfo {author} {\bibfnamefont {S.}~\bibnamefont {Miki}}, \bibinfo {author} {\bibfnamefont {T.}~\bibnamefont {Yamashita}}, \bibinfo {author} {\bibfnamefont {Z.}~\bibnamefont {Wang}}, \bibinfo {author} {\bibfnamefont {A.}~\bibnamefont {Tanaka}}, \bibinfo {author} {\bibfnamefont {K.}~\bibnamefont {Yoshino}}, \bibinfo {author} {\bibfnamefont {Y.}~\bibnamefont {Nambu}}, \bibinfo {author} {\bibfnamefont {S.}~\bibnamefont {Takahashi}}, \bibinfo {author} {\bibfnamefont {A.}~\bibnamefont {Tajima}}, \bibinfo {author} {\bibfnamefont {A.}~\bibnamefont {Tomita}}, \bibinfo {author} {\bibfnamefont {T.}~\bibnamefont {Domeki}},
  \bibinfo {author} {\bibfnamefont {T.}~\bibnamefont {Hasegawa}}, \bibinfo {author} {\bibfnamefont {Y.}~\bibnamefont {Sakai}}, \bibinfo {author} {\bibfnamefont {H.}~\bibnamefont {Kobayashi}}, \bibinfo {author} {\bibfnamefont {T.}~\bibnamefont {Asai}}, \bibinfo {author} {\bibfnamefont {K.}~\bibnamefont {Shimizu}}, \bibinfo {author} {\bibfnamefont {T.}~\bibnamefont {Tokura}}, \bibinfo {author} {\bibfnamefont {T.}~\bibnamefont {Tsurumaru}}, \bibinfo {author} {\bibfnamefont {M.}~\bibnamefont {Matsui}}, \bibinfo {author} {\bibfnamefont {T.}~\bibnamefont {Honjo}}, \bibinfo {author} {\bibfnamefont {K.}~\bibnamefont {Tamaki}}, \bibinfo {author} {\bibfnamefont {H.}~\bibnamefont {Takesue}}, \bibinfo {author} {\bibfnamefont {Y.}~\bibnamefont {Tokura}}, \bibinfo {author} {\bibfnamefont {J.~F.}\ \bibnamefont {Dynes}}, \bibinfo {author} {\bibfnamefont {A.~R.}\ \bibnamefont {Dixon}}, \bibinfo {author} {\bibfnamefont {A.~W.}\ \bibnamefont {Sharpe}}, \bibinfo {author} {\bibfnamefont {Z.~L.}\ \bibnamefont {Yuan}}, \bibinfo
  {author} {\bibfnamefont {A.~J.}\ \bibnamefont {Shields}}, \bibinfo {author} {\bibfnamefont {S.}~\bibnamefont {Uchikoga}}, \bibinfo {author} {\bibfnamefont {M.}~\bibnamefont {Legr\'{e}}}, \bibinfo {author} {\bibfnamefont {S.}~\bibnamefont {Robyr}}, \bibinfo {author} {\bibfnamefont {P.}~\bibnamefont {Trinkler}}, \bibinfo {author} {\bibfnamefont {L.}~\bibnamefont {Monat}}, \bibinfo {author} {\bibfnamefont {J.-B.}\ \bibnamefont {Page}}, \bibinfo {author} {\bibfnamefont {G.}~\bibnamefont {Ribordy}}, \bibinfo {author} {\bibfnamefont {A.}~\bibnamefont {Poppe}}, \bibinfo {author} {\bibfnamefont {A.}~\bibnamefont {Allacher}}, \bibinfo {author} {\bibfnamefont {O.}~\bibnamefont {Maurhart}}, \bibinfo {author} {\bibfnamefont {T.}~\bibnamefont {L\"{a}nger}}, \bibinfo {author} {\bibfnamefont {M.}~\bibnamefont {Peev}},\ and\ \bibinfo {author} {\bibfnamefont {A.}~\bibnamefont {Zeilinger}},\ }\bibfield  {title} {\bibinfo {title} {Field test of quantum key distribution in the tokyo qkd network},\ }\href
  {https://doi.org/10.1364/OE.19.010387} {\bibfield  {journal} {\bibinfo  {journal} {Opt. Express}\ }\textbf {\bibinfo {volume} {19}},\ \bibinfo {pages} {10387} (\bibinfo {year} {2011})}\BibitemShut {NoStop}%
\bibitem [{\citenamefont {Dynes}\ \emph {et~al.}(2012)\citenamefont {Dynes}, \citenamefont {Choi}, \citenamefont {Sharpe}, \citenamefont {Dixon}, \citenamefont {Yuan}, \citenamefont {Fujiwara}, \citenamefont {Sasaki},\ and\ \citenamefont {Shields}}]{Dynes:12}%
  \BibitemOpen
  \bibfield  {author} {\bibinfo {author} {\bibfnamefont {J.~F.}\ \bibnamefont {Dynes}}, \bibinfo {author} {\bibfnamefont {I.}~\bibnamefont {Choi}}, \bibinfo {author} {\bibfnamefont {A.~W.}\ \bibnamefont {Sharpe}}, \bibinfo {author} {\bibfnamefont {A.~R.}\ \bibnamefont {Dixon}}, \bibinfo {author} {\bibfnamefont {Z.~L.}\ \bibnamefont {Yuan}}, \bibinfo {author} {\bibfnamefont {M.}~\bibnamefont {Fujiwara}}, \bibinfo {author} {\bibfnamefont {M.}~\bibnamefont {Sasaki}},\ and\ \bibinfo {author} {\bibfnamefont {A.~J.}\ \bibnamefont {Shields}},\ }\bibfield  {title} {\bibinfo {title} {Stability of high bit rate quantum key distribution on installed fiber},\ }\href {https://doi.org/10.1364/OE.20.016339} {\bibfield  {journal} {\bibinfo  {journal} {Opt. Express}\ }\textbf {\bibinfo {volume} {20}},\ \bibinfo {pages} {16339} (\bibinfo {year} {2012})}\BibitemShut {NoStop}%
\bibitem [{\citenamefont {Dixon}\ \emph {et~al.}(2015)\citenamefont {Dixon}, \citenamefont {Dynes}, \citenamefont {Lucamarini}, \citenamefont {Fr\"{o}hlich}, \citenamefont {Sharpe}, \citenamefont {Plews}, \citenamefont {Tam}, \citenamefont {Yuan}, \citenamefont {Tanizawa}, \citenamefont {Sato}, \citenamefont {Kawamura}, \citenamefont {Fujiwara}, \citenamefont {Sasaki},\ and\ \citenamefont {Shields}}]{Dixon:15}%
  \BibitemOpen
  \bibfield  {author} {\bibinfo {author} {\bibfnamefont {A.~R.}\ \bibnamefont {Dixon}}, \bibinfo {author} {\bibfnamefont {J.~F.}\ \bibnamefont {Dynes}}, \bibinfo {author} {\bibfnamefont {M.}~\bibnamefont {Lucamarini}}, \bibinfo {author} {\bibfnamefont {B.}~\bibnamefont {Fr\"{o}hlich}}, \bibinfo {author} {\bibfnamefont {A.~W.}\ \bibnamefont {Sharpe}}, \bibinfo {author} {\bibfnamefont {A.}~\bibnamefont {Plews}}, \bibinfo {author} {\bibfnamefont {S.}~\bibnamefont {Tam}}, \bibinfo {author} {\bibfnamefont {Z.~L.}\ \bibnamefont {Yuan}}, \bibinfo {author} {\bibfnamefont {Y.}~\bibnamefont {Tanizawa}}, \bibinfo {author} {\bibfnamefont {H.}~\bibnamefont {Sato}}, \bibinfo {author} {\bibfnamefont {S.}~\bibnamefont {Kawamura}}, \bibinfo {author} {\bibfnamefont {M.}~\bibnamefont {Fujiwara}}, \bibinfo {author} {\bibfnamefont {M.}~\bibnamefont {Sasaki}},\ and\ \bibinfo {author} {\bibfnamefont {A.~J.}\ \bibnamefont {Shields}},\ }\bibfield  {title} {\bibinfo {title} {High speed prototype quantum key distribution system and
  long term field trial},\ }\href {https://doi.org/10.1364/OE.23.007583} {\bibfield  {journal} {\bibinfo  {journal} {Opt. Express}\ }\textbf {\bibinfo {volume} {23}},\ \bibinfo {pages} {7583} (\bibinfo {year} {2015})}\BibitemShut {NoStop}%
\bibitem [{\citenamefont {Santagiustina}\ \emph {et~al.}(2024)\citenamefont {Santagiustina}, \citenamefont {Agnesi}, \citenamefont {Alarc\'{o}n}, \citenamefont {Cabello}, \citenamefont {Xavier}, \citenamefont {Villoresi},\ and\ \citenamefont {Vallone}}]{Santagiustina:24}%
  \BibitemOpen
  \bibfield  {author} {\bibinfo {author} {\bibfnamefont {F.~B.~L.}\ \bibnamefont {Santagiustina}}, \bibinfo {author} {\bibfnamefont {C.}~\bibnamefont {Agnesi}}, \bibinfo {author} {\bibfnamefont {A.}~\bibnamefont {Alarc\'{o}n}}, \bibinfo {author} {\bibfnamefont {A.}~\bibnamefont {Cabello}}, \bibinfo {author} {\bibfnamefont {G.~B.}\ \bibnamefont {Xavier}}, \bibinfo {author} {\bibfnamefont {P.}~\bibnamefont {Villoresi}},\ and\ \bibinfo {author} {\bibfnamefont {G.}~\bibnamefont {Vallone}},\ }\bibfield  {title} {\bibinfo {title} {Experimental post-selection loophole-free time-bin and energy-time nonlocality with integrated photonics},\ }\href {https://doi.org/10.1364/OPTICA.499247} {\bibfield  {journal} {\bibinfo  {journal} {Optica}\ }\textbf {\bibinfo {volume} {11}},\ \bibinfo {pages} {498} (\bibinfo {year} {2024})}\BibitemShut {NoStop}%
\end{thebibliography}%

\end{document}